\documentclass[twocolumn]{openjournal}

\usepackage{hyperref}
\usepackage{verbatim}
\graphicspath{{./}{figures/}}

\newcommand{\nObjects}{2{,}569}
\newcommand{\nClusters}{1{,}588}
\newcommand{\nGalaxies}{981}
\newcommand{\nObsIDs}{5{,}505}
\newcommand{\nMultiObsID}{744}
\newcommand{\maxObsIDsPerObject}{135}
\newcommand{\nWithRedshift}{2{,}110}
\newcommand{\medianRedshift}{0.064}
\newcommand{\maxRedshift}{3.79}
\newcommand{\totalExposureMs}{138.8}
\newcommand{\medianExposureKs}{21.8}
\newcommand{\maxExposureMs}{3.32}
\newcommand{\nAboveHundredKs}{295}

\newcommand{\nCentersPeak}{1{,}533}
\newcommand{\nCentersFallback}{1{,}036}
\newcommand{\nCentersFallbackClusters}{281}
\newcommand{\nCentersFallbackGalaxies}{755}
\newcommand{\medianCenterOffset}{18}

\newcommand{\lowCenterOffset}{8}
\newcommand{\highCenterOffset}{66}

\newcommand{\nCenterBeyondTwoArcmin}{136}
\newcommand{\nEmbeddings}{2{,}569}
\newcommand{\nNeighbors}{10}
\newcommand{\medianTopSimilarity}{0.967}
\newcommand{\medianSimilarity}{0.955}
\newcommand{\nEmbeddedClusters}{1{,}588}
\newcommand{\neighborDlogZ}{0.42}
\newcommand{\randomDlogZ}{0.92}
\newcommand{\neighborDlogZClusters}{0.26}
\newcommand{\randomDlogZClusters}{0.57}
\newcommand{\nAcceptMatched}{229}
\newcommand{\nAcceptTotal}{240}
\newcommand{\nAcceptCenterMatched}{235}
\newcommand{\acceptCenterOffset}{8}
\newcommand{\acceptCenterLow}{4}
\newcommand{\acceptCenterHigh}{15}

\newcommand{\nAcceptPairs}{1{,}105}
\newcommand{\acceptCoolCorePercent}{45}
\newcommand{\ccAgreementPercent}{73}
\newcommand{\ccNullPercent}{51}
\newcommand{\ccSignificance}{11}
\newcommand{\ccGivenCCPercent}{64}
\newcommand{\ccGivenNCCPercent}{19}
\newcommand{\nPermutations}{5{,}000}

\newcommand{\nValSources}{275}
\newcommand{\nValObjects}{237}
\newcommand{\valBroadRatio}{0.971}
\newcommand{\valBroadLow}{0.94}
\newcommand{\valBroadHigh}{1.01}
\newcommand{\valSoftRatio}{0.803}
\newcommand{\valHardRatio}{0.956}
\newcommand{\valSoftLow}{0.69}
\newcommand{\valSoftHigh}{0.87}
\newcommand{\valHardLow}{0.90}
\newcommand{\valHardHigh}{1.02}
\newcommand{\valSoftEarly}{0.83}
\newcommand{\valSoftLate}{0.70}

\newcommand{\nBkgObjects}{300}
\newcommand{\nBkgQuiet}{134}
\newcommand{\bkgQuietMedian}{2.5}
\newcommand{\bkgQuietLow}{$-$1}
\newcommand{\bkgQuietHigh}{12}
\newcommand{\bkgQuietWithinTen}{81}

\newcommand{\pctNoBkgModel}{6}

\defcitealias{planck_collaboration_planck_2020}{Planck Collaboration 2020}
\defcitealias{collaboration_astropy_2022}{Astropy Collaboration 2022}

\begin{document}

\title{LEMUR: A Searchable Web Archive of Deep Chandra Images of Galaxies and Galaxy Clusters}

\author{Carter Lee Rhea}
\affiliation{Dragonfly Focused Research Organization, 150 Washington Avenue, Santa Fe, NM 87501, USA}
\affiliation{Centre de recherche en astrophysique du Qu\'ebec (Astroquébec)}
\email{carterrhea93@proton.me}

\author{Julie Hlavacek-Larrondo}
\affiliation{D\'epartement de physique, Universit\'e de Montr\'eal, C.P. 6128 Succ. Centre-ville, Montr\'eal H3C 3J7, Canada}

\author{Marine Prunier}
\affiliation{MIT Kavli Institute for Astrophysics and Space Research, Cambridge, MA 02139, USA}

\author{Hyunseop Choi}
\affiliation{Department of Astronomy, University of Michigan, 1085 S. University Ave., Ann Arbor, MI 48109, USA}

\author{Megan Donahue}
\affiliation{Michigan State University, Department of Physics \& Astronomy, 567 Wilson Road, East Lansing, MI 48824, USA}

\author{Christian Thibeault}
\affiliation{D\'epartement de physique, Universit\'e de Montr\'eal, C.P. 6128 Succ. Centre-ville, Montr\'eal H3C 3J7, Canada}

\begin{abstract}
We present LEMUR, a publicly accessible archive of uniformly reduced, deep \textit{Chandra} images of \nObjects\ galaxies and galaxy clusters (including groups), comprising \nObsIDs\ ACIS observations and \totalExposureMs\,Ms of exposure. By combining the publicly available imaging observations of each target, we provide background-subtracted, exposure-corrected images in broad, soft and hard bands at the native ACIS pixel scale of $0.492''$. Our primary goal is to make the wealth of archival \textit{Chandra} observations accessible to researchers without requiring specialized expertise in X-ray data reduction.
Each observation is reprocessed with CIAO, filtered for background flares and reduced with settings chosen to preserve faint diffuse emission; the merged images are exposure corrected and background subtracted using the ACIS blank-sky backgrounds in a broad (0.5--7.0\,keV), a soft (0.5--2.0\,keV) and a hard (2.0--7.0\,keV) band. For every object, we measure an X-ray center at the peak of its diffuse emission, and we provide a morphological similarity search based on embeddings from the AstroCLIP foundation model, in which nearest neighbors share cool-core status in \ccAgreementPercent\% of matched cluster pairs, compared with \ccNullPercent\% by chance.
Photometry of \nValSources\ point sources agrees with the Chandra Source Catalog to within 3\% in the broad band. The images, with their counts images, background models and exposure maps, and quick-look stamps are available through a searchable web archive at \url{https://lemur-archive.lemur-xray.workers.dev/}, the catalog tables are archived on Zenodo (\href{https://doi.org/10.5281/zenodo.24351915}{doi:10.5281/zenodo.24351915}), and all code is open source at \url{https://github.com/crhea93/Lemur}.

\end{abstract}


\section{Introduction} \label{sec:intro}
Over the past 25 years, Chandra has been instrumental in our understanding of the hot, diffuse gas in galaxies and galaxy clusters (\citealt{white_high_2009}). Through using the Chandra archive, astronomers have discovered the presence of cold fronts in the intra-cluster medium (ICM) of galaxy clusters (\citealt{owers_high_2009}), studied the statistics of X-shaped radio galaxies (\citealt{hodges-kluck_chandra_2010}), and verified the existence of hot, X-ray emitting gas in dwarf galaxies (\citealt{ott_chandra_2005,pandya_comprehensive_2016}). While these studies have advanced our understanding of X-ray emitting sources, the full potential of the Chandra archive remains unrealized, in part because few systematically reduced, merged images of these objects are available.

Chandra archival data is widely available through a web-based data portal, the \href{https://cda.harvard.edu/chaser/}{Chandra Data Archive} (CDA).
The portal offers calibrated data products for each archival Chandra observation, including level-1 and level-2 processed event files, exposure files, and observational reports, but not reduced images that combine the observations of a target.
Other archives, such as the Archive of Chandra Cluster Entropy Profile Tables  \citep[ACCEPT]{cavagnolo_intracluster_2009} and its recent successor ACCEPT 2.0 \citep{donahue_accept_2026}, provide uniformly derived thermodynamic properties of galaxy clusters. ACCEPT 2.0 expands the original sample to 532 clusters, providing radial density, temperature, and entropy profiles, along with integrated X-ray properties. LEMUR\footnote{LEMUR is not an acronym: the archive is named after the first author's cat, Lemur, who in turn is named after her ringed tail.} complements these efforts by focusing on uniformly processed, deep, merged X-ray imaging products for a broader sample of galaxies and galaxy clusters.
%other archives such as ACCEPT (Archive of Chandra Cluster Entropy Profiles Table; \citealt{cavagnolo_intracluster_2009}) provide computed spectroscopic and photometric values such as entropy profiles, temperature profiles, and bolometric luminosities for active galaxies and galaxy clusters. 
The CXO also provides tools for a ``quick look'' at observations (Chandra Fast Image\footnote{\url{https://cda.harvard.edu/pop/dispatchQuick}}) and cleaned data from individual observations (Chandra Footprint Service\footnote{\url{https://cxcfps.cfa.harvard.edu/cda/footprint/cdaview.html}}); however, no service provides merged images of the many targets observed more than once (\nMultiObsID\ of the \nObjects\ objects in LEMUR).

Several other catalogs draw on the X-ray archives, but none provides deep, merged images of individual extended sources. The Chandra Source Catalog \citep[CSC]{evans_chandra_2010,evans_chandra_2024} combines overlapping Chandra observations to detect sources and reports their positions, fluxes and spectral properties; its stacked images combine only observations whose pointings lie within about an arcminute of each other, and its detection is designed for point-like and compact sources rather than emission extending over arcminutes. The Chandra Galaxy Atlas \citep{kim_chandra_2019} provides uniformly reduced, merged images and spatially resolved spectral maps of 70 early-type galaxies. The first eROSITA all-sky survey \citep[eRASS1;][]{merloni_srgerosita_2024} and its catalog of galaxy clusters \citep{bulbul_srgerosita_2024} cover half of the sky uniformly, but with exposures of a few hundred seconds and an angular resolution of roughly half an arcminute, far shallower and coarser than a deep Chandra pointing. The XMM Cluster Survey \citep[XCS]{romer_serendipitous_2001,mehrtens_xmm_2012} searches the XMM-Newton archive for serendipitously detected clusters and measures their properties, while ACCEPT and ACCEPT 2.0 provide entropy and temperature profiles for a sample of well-observed clusters. LEMUR is complementary to these efforts. For every galaxy and galaxy cluster observed with Chandra ACIS under the CDA's "normal galaxies" and "clusters of galaxies" science categories, it combines all available observations into a single, background-subtracted and exposure-corrected image reduced to preserve faint diffuse emission, measures the position of that emission, and makes the result searchable, including by morphological similarity to other objects in the catalog.

In this manuscript, we present an archive of deep, merged images of galaxies and galaxy clusters, reduced with a pipeline optimized for the detection of faint diffuse emission.
The primary aim of this catalog is to make available the deepest Chandra reduction of galaxies and galaxy clusters for non X-ray astronomers and X-ray astronomers alike; therefore, it improves the dissemination of Chandra data products. To make the final data products widely accessible, we have created a modern web-based catalog and provide all code for the construction of the data products and website at \url{https://github.com/crhea93/Lemur}.

In Section~\ref{sec:methods}, we describe how the data were selected and grouped, the main components of the reduction code, how we locate each object, the similarity search, and the scaffolding of the website. In Section~\ref{sec:results}, we summarize the catalog, validate its images and compare two of its reductions with published analyses. In Section~\ref{sec:conclusion}, we discuss the implications of this catalog and future improvements.

\section{Methods} \label{sec:methods}
\subsection{Data}

The data were downloaded directly from the CDA by querying all ACIS observations whose initial submission was categorized as either a ``normal galaxy" or a ``cluster of galaxies". We note that while galaxy groups is not a distinct submission category, they typically fall under the "cluster of galaxies" category. ACIS has $0.492''$ pixels \citep{garmire_advanced_2003}, comparable to the on-axis point-spread function of the telescope, which broadens to several arcseconds a few arcminutes off axis. The search, last updated on 2026 November 2, returned 2,107 and 4,013 archived observations of normal galaxies and galaxy clusters, respectively. We exclude the 276 imaging observations made for instrument calibration (proposal type CAL, 4.6\,Ms in total), most of them short, repeated pointings of calibration targets such as Abell\,1795 and the Perseus cluster. We also removed observations of blank sky and three targets that are mislabeled as galaxy or galaxy cluster objects (\textit{NGC3221Off-field}, \textit{OffsetTDE2} and \textit{PFF2004gc100}), as well as one field of the Galactic Bulge Latitude Survey, which consists of 36 short ACIS-I pointings tiled over $\sim2.5^\circ$ of the bulge rather than observations of a single extended source.

Because LEMUR provides images, we keep only observations that produce one. We removed the 68 observations taken with the High Energy Transmission Grating in place, and the four read out in continuous-clocking mode, which have no two-dimensional image: three of M87 and the only observation of the Sagittarius dwarf spheroidal galaxy (ObsID 3927). Seventy-eight observations, 76 of M31 and 2 of M32, were taken in interleaved (alternating exposure) mode, which records events with two frame times; we use the events of the longer frame time, which hold 97\% of the exposure. One archived observation (ObsID 963 of NGC\,4406) contains no events and contributes nothing to its object's image.

In order to group observations by target, we combined all observations that shared a target name (e.g., Perseus). Since several targets go by different names, we additionally combined ObsIDs whose target names resolved to the same object in NED (e.g., Perseus, NGC1275, PerseusNE) or differ only in the catalog prefix or formatting (e.g., A2029 and Abell2029). Finally, we combined all targets whose centers were within 10 arcseconds of each other. Grouping by name can also attach an observation to the wrong object, for example when two names normalize to the same string: ObsID 3233 of 3C\,28.0 had been grouped with 3C\,280. We therefore inspected every object whose observations are pointed more than $30'$ apart. Most are genuine mosaics of nearby objects, but six observations had been grouped with the wrong object; we removed them.

This grouping defines the objects. To make sure that every observation in the selection is used, we then compared the grouped catalog with the archive and assigned each remaining observation. An observation joins an object when its target name matches one of the object's names and it lies within $30'$ of the object's pointings, or when it lies within $3.5'$ of one of those pointings, in which case it images the same field under another name. Pointings whose names mark them as offset fields of an object, such as ``Perseus cluster N'', ``Abell 1795 outskirts'' and ``A85 North'' (199 observations, 4.5\,Ms in total), are neither merged into that object nor added as new objects; a few offset pointings that the grouping above had already made into separate objects (for example A2259~SE, SW and NW) remain in the catalog as such. All other targets form new objects.

For three nearby objects, the archive also contains pointings far from the region covered by the deep observations: four 10\,ks pointings 57--111$'$ from the core of the Centaurus cluster, three pointings $\sim75'$ from the nucleus of M31, and 20 pointings 57--82$'$ from NGC\,1275 in the Perseus cluster, among them a ring of pointings near $r_{200}$. Imaging these at the native resolution would enlarge the merged image of each object several times over while adding little exposure on the object itself, so for these three objects we merge only the observations within $30'$ of the core: 15 of 19 ObsIDs for Centaurus, 135 of 138 for M31 and 121 of 141 for Perseus (3.4 of 3.8\,Ms). The excluded observations remain available from the CDA.

This resulted in a total of \nGalaxies\ galaxies and \nClusters\ clusters, built from \nObsIDs\ unique ObsIDs; their sky and redshift distributions are shown in Figure~\ref{fig:catalog} and discussed in Section~\ref{sec:stats}.
Object positions are described in Section~\ref{sec:centers}. The redshift of each object was resolved from NED or the CDS; in the case where no redshift was available due to lack of a measurement in the literature, we adopt a value of \textit{N/A}.

\subsection{Reduction and Analysis}

\subsubsection{Cleaning and Merging} \label{sec:cleaning}
The reduction pipeline consists of three main components: reprocessing individual ObsIDs, subtracting the background, and merging the ObsIDs for a single target. The first part of the pipeline begins with the determination of a good time interval (GTI). To do so, we use as our background CCD the chip of the observation with the fewest total counts, which is normally the chip least affected by the target. We then run \texttt{vtpdetect} \citep{ebeling_detecting_1993} on this chip to flag and remove all point-like sources, leaving behind only diffuse emission or background emission for the light curve; the catalog images themselves keep all point sources. In the case of galaxies and galaxy clusters, the X-ray emission of diffuse ionized gas varies on timescales considerably longer than a typical Chandra exposure; therefore, even if the target lies on the CCD, it will not affect the computation of the good time interval. For the cleaned background CCD, we construct a light curve in 200\,s bins and remove the time bins whose count rate lies more than $3\sigma$ from the mean (\texttt{lc\_sigma\_clip}), which captures strong background flares. If the CCD does not contain a sufficient count rate to compute the GTI robustly, or if the clipped GTI retains less than 10\% of the observation's good time, we fall back to the GTI provided in the level-2 event file. Because the background CCD is faint, and often a front-illuminated chip, which is less sensitive to soft-proton flares than the back-illuminated chips, this filtering is conservative: it removes strong flares but can leave moderate ones in the data (see Section~\ref{sec:showcase}). Observations taken in several segments (OBIs) are cleaned segment by segment, each with its own light curve and calibration files, and the segments are combined before merging. Subsequently, we remove streaks in the image caused by CCD readout issues using the \texttt{CIAO destreak} tool. We then run \texttt{acis\_build\_badpix} to flag known bad pixels or those with bad bias values. Then, we run \texttt{acis\_find\_afterglow} to detect cosmic ray afterglows and hot pixels. The result is then combined with a second run of \texttt{acis\_build\_badpix} to remove those pixels. We then pass the cleaned files to \texttt{acis\_process\_events} with \texttt{eventdef=)stdlev1} and \texttt{check\_vf\_pha=True} in order to detect faint emission at the expense of potentially misclassifying point sources as cosmic ray hits. The processed event file is filtered by ``good'' grades as classified by \texttt{CIAO}. Finally, the GTI is applied to create a final processed observation. We note that this follows the standard \texttt{CIAO} reprocessing procedure, \texttt{chandra\_repro}, closely with small departures to optimize for faint emission. All observations were reduced with CIAO 4.18.0 \citep{fruscione_ciao_2006}, CALDB 4.12.4 and version 4.9.7 of the ACIS blank-sky background files.

For each ObsID (and each segment of a multi-segment ObsID), we then create a background event file with the \texttt{blanksky} command, which selects the matching ACIS blank-sky background from the calibration database, reprojects it with the aspect solution of the observation, and scales it chip by chip to the observed 9--12\,keV particle background. The scaling uses all events of each chip in that band; the target contributes little at these energies, where the telescope has almost no effective area. The blank-sky events are filtered for VFAINT mode in the same way as the observation, and for a chip without events at 9--12\,keV the background is scaled by exposure time instead. For some observations, the calibration database has no blank-sky background for one or more of the chips; we build the background from the chips that have one, and the remaining chips are exposure corrected but not background subtracted (about \pctNoBkgModel\% of the objects; Section~\ref{sec:validation}). The cleaned event files of all ObsIDs of an object are merged and exposure corrected in the 0.5--7.0\,keV band with \texttt{merge\_obs} at the native ACIS pixel scale ($0.492''$), with the exposure map computed at an effective energy of 2.3\,keV; objects observed only once are imaged with \texttt{fluximage} at the same scale. To subtract the background, we reproject each blank-sky event file onto the sky frame of the merged image with \texttt{reproject\_events}, bin it onto the merged image grid in the same band, and apply the per-chip scaling. The summed background counts $B$ are subtracted from the merged counts image $C$ before dividing by the merged exposure map $E$, so that the catalog image is $F=(C-B)/E$ on every exposed pixel.

We make images in a soft (0.5--2.0\,keV) and a hard (2.0--7.0\,keV) band in the same way: \texttt{merge\_obs} is run on the same cleaned event files in each band, with exposure maps computed at effective energies of 1.25 and 3.8\,keV, and the blank-sky events are filtered to the band before binning. The per-chip scaling, measured at 9--12\,keV, is the same in every band.

For each object and band, we provide the background-subtracted, exposure-corrected image $F$ (in photons\,cm$^{-2}$\,s$^{-1}$ per pixel; since each exposure map is computed at a single energy, $F$ is an approximate photon flux, see Section~\ref{sec:validation}), the counts image $C$ and the background model $B$ (both in counts), together with the exposure map $E$ of the broad band. We do not provide separate soft- and hard-band exposure maps: they follow from $E=(C-B)/F$ only where $F\neq0$, which leaves many exposed pixels of low-count images undefined, so users who need them should recompute them with \texttt{merge\_obs} or \texttt{fluximage}. We also provide a postage stamp PNG of each of the three images. To make it, we bin images larger than 2048 pixels on a side down to that size, convolve the image with a Gaussian kernel ($\sigma=2$ pixels), and normalize it with an \texttt{arcsinh} stretch function with $a=0.08$\footnote{\href{https://docs.astropy.org/en/stable/api/astropy.visualization.AsinhStretch.html}{\texttt{astropy.AsinhStretch}}}.

\subsection{Object Positions and X-ray Centers} \label{sec:centers}

We report two positions for every object in the catalog. The \textit{target position} is the pointing requested by the observer, taken from the \texttt{RA\_TARG} and \texttt{DEC\_TARG} keywords of the merged image. If these keywords are missing, we resolve the target name with NED and then with the CDS, and as a last resort we use the nominal pointing (\texttt{RA\_NOM}, \texttt{DEC\_NOM}) of the observation. The target position records what the observer intended to study, but it need not coincide with the X-ray emission: observers often offset the aimpoint, and for extended or merging systems the target may be an optical or SZ position. We therefore also measure the \textit{X-ray center}, defined as the peak of the diffuse emission in the merged, exposure-corrected 0.5--7.0\,keV image. We use this center wherever the catalog needs a single reference point, including the cutouts for the similarity search (Section~\ref{sec:astroclip}).

We do not use a flux-weighted centroid of the full field. For a multi-chip ACIS image, the centroid is dominated by the background over the whole array and is pulled toward the middle of the field, often arcminutes away from the source.

The peak finder is designed to find extended emission on scales of tens of arcseconds while ignoring point sources, chip edges and chip-to-chip differences in background. We first bin the image to a working resolution of $3.9''$ (eight ACIS pixels) and build a coverage mask from the exposure map, excluding pixels whose exposure is below 10\% of the 90th percentile. We erode the mask by $30''$ so that chip edges cannot produce a peak. Point sources are then flagged as compact excesses: pixels where the image smoothed on a $4''$ scale exceeds the image smoothed on a $20''$ scale by more than five robust standard deviations (estimated from the median absolute deviation) and by more than a factor of two. We mask these pixels after dilating them by $8''$.

All smoothing uses a mask-aware Gaussian filter, which convolves the masked image and the mask with the same kernel and divides the two. This prevents masked regions and field edges from lowering the smoothed surface brightness nearby. We locate the emission on a band-pass image, formed as the image smoothed with a $\sigma=30''$ Gaussian minus the image smoothed with a $\sigma=120''$ Gaussian. The band-pass keeps structure on cluster-core and galaxy-halo scales and removes the large-scale background, so a uniformly brighter chip (for example the back-illuminated ACIS-S3) does not win.

We search for the peak within $4'$ of the target position for clusters and within $1'$ for galaxies. Cluster targets are often optical or SZ positions that can lie arcminutes from the X-ray peak, whereas galaxy targets are optical positions accurate to arcseconds, so emission arcminutes from a galaxy belongs to another source, such as a knot in a spiral arm or the edge of a brighter chip. Brighter peaks elsewhere in the field likewise belong to other sources, such as a foreground cluster or the parent cluster of an offset pointing, so we do not consider them. If the brightest band-pass value in the search region lies on its edge, the emission continues outward: for clusters we follow the band-pass image uphill to its maximum, and for galaxies we treat the emission as unrelated to the target. We accept the peak only if it exceeds the median of the band-pass image by at least five robust standard deviations; in fields without extended emission, the brightest band-pass value within $4'$ of the target reaches two to four. Finally, we refine the peak by smoothing the image with an $8''$ Gaussian and taking the maximum within $45''$ of the band-pass peak. The resulting center is quantized to the $3.9''$ working pixel, which is adequate for a catalog reference position but is not a substitute for a dedicated centroid or surface brightness fit. If no significant peak is found, we adopt the target position as the X-ray center and flag it in the catalog. This is the case for \nCentersFallback\ objects: \nCentersFallbackGalaxies\ galaxies without detectable extended emission and \nCentersFallbackClusters\ clusters, among them distant clusters, compact groups and pointings at the outskirts of clusters.

\begin{figure}
    \centering
    \includegraphics[width=0.95\linewidth]{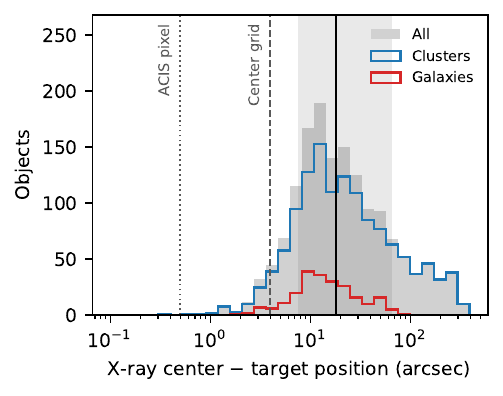}
    \caption{Separation between the X-ray center and the target position for the objects with a significant X-ray peak (gray), and separately for clusters and galaxies. The solid line marks the median and the shaded band the 16th--84th percentiles of all objects shown. The dotted and dashed lines mark the ACIS pixel ($0.49''$) and the $3.9''$ grid on which the centers are measured.}
    \label{fig:center_offsets}
\end{figure}

Figure~\ref{fig:center_offsets} shows the separation between the X-ray center and the target position for the \nCentersPeak\ objects with a significant peak. The median separation is $\medianCenterOffset''$, and 68\% of the objects lie between $\lowCenterOffset''$ and $\highCenterOffset''$ from their target. These separations are much larger than the ACIS pixel and the $3.9''$ grid on which the centers are measured, so they reflect the target positions rather than the measurement. For clusters they follow the positional accuracy of the catalog that supplied the target, which we identify from the prefix of the object name: the median separation is about $10''$ for MACS and ACT clusters, $15$--$20''$ for ROSAT and SPT clusters, $25''$ for Abell clusters, whose positions are optical galaxy-density centers, and $60''$ for Planck clusters, whose positions come from a beam of several arcminutes. Physical offsets between the X-ray peak and the optical or SZ center of merging clusters add to these. The \nCenterBeyondTwoArcmin\ objects more than $2'$ from their target are all clusters: clusters with SZ or optical target positions, merging clusters, and pointings at the outskirts of a bright cluster, for which the X-ray center follows the emission to its peak.

The X-ray centers agree with independent measurements. For \nAcceptCenterMatched\ of the \nAcceptTotal\ clusters in ACCEPT \citep{cavagnolo_intracluster_2009}, the nearest LEMUR X-ray center lies within $2'$ of the center adopted by ACCEPT, at a median separation of $\acceptCenterOffset''$ (68\% between $\acceptCenterLow''$ and $\acceptCenterHigh''$), comparable to the $3.9''$ grid on which our centers are measured.

\subsection{Similarity Search with AstroCLIP} \label{sec:astroclip}
To let users find objects that look alike, we embed each merged image with the image encoder of AstroCLIP \citep{parker_astroclip_2024} and return, for every object, its nearest neighbors in the embedding space.
AstroCLIP is a foundation model for galaxies: its image encoder is a vision transformer pretrained with self-supervised learning on $g$, $r$, $z$ cutouts from the DESI Legacy Imaging Surveys \citep{dey_overview_2019} and then aligned with DESI spectra through a contrastive (CLIP) objective. We use the publicly released checkpoint without retraining. Chandra images of diffuse X-ray emission are therefore far outside the model's training distribution; we use the encoder as a generic descriptor of image morphology, and the preprocessing below is designed to present the X-ray images in a form close to what the encoder expects.

For each object, we cut out a $512\times512$ native ACIS pixel ($4.2'\times4.2'$) region of the merged, background-subtracted, exposure-corrected 0.5--7.0\,keV image, centered on the X-ray center (Section~\ref{sec:centers}). Regions that fall off the detector are set to zero, so the X-ray center always sits at the middle of the cutout. We first resample the cutout onto a grid of $3.9''$ pixels ($64\times64$ pixels) with a flux-conserving area-weighted resampling. We then smooth with a Gaussian of $\sigma=1.5$ grid pixels ($5.9''$), which turns the sparse single-photon pixels of low-count observations into visible diffuse emission, and stretch the image into $[0,1]$: we clip between the 5th and 99.7th percentiles of the positive pixels and apply the same $\mathrm{arcsinh}$ stretch ($a=0.08$) used for the postage stamps. The per-image normalization removes differences in exposure time and flux, so the embedding responds to the shape of the emission rather than its brightness. Finally, the image is resampled to the $144\times144$ input size of the encoder and replicated into the three input channels. \citet{karam_cool_2026} applied the same encoder with a similar preprocessing to mock X-ray images of simulated clusters to predict their cooling times.

The encoder maps each image to a 1024-dimensional vector, which we normalize to unit length. We measure similarity between two objects as the cosine similarity of their embeddings and store the \nNeighbors\ most similar objects for each of the \nEmbeddings\ embedded objects (\nEmbeddedClusters\ galaxy clusters and the rest galaxies); the neighbors are shown on each object's page of the online catalog. Because all inputs are out of distribution, the embeddings occupy a narrow region of the space: the median similarity between an object and its nearest neighbor is \medianTopSimilarity, and \medianSimilarity\ over all \nNeighbors\ neighbors. The absolute similarity values should therefore not be interpreted on their own; only the ranking of neighbors is meaningful.

\begin{figure*}
    \centering
    \includegraphics[width=0.95\linewidth]{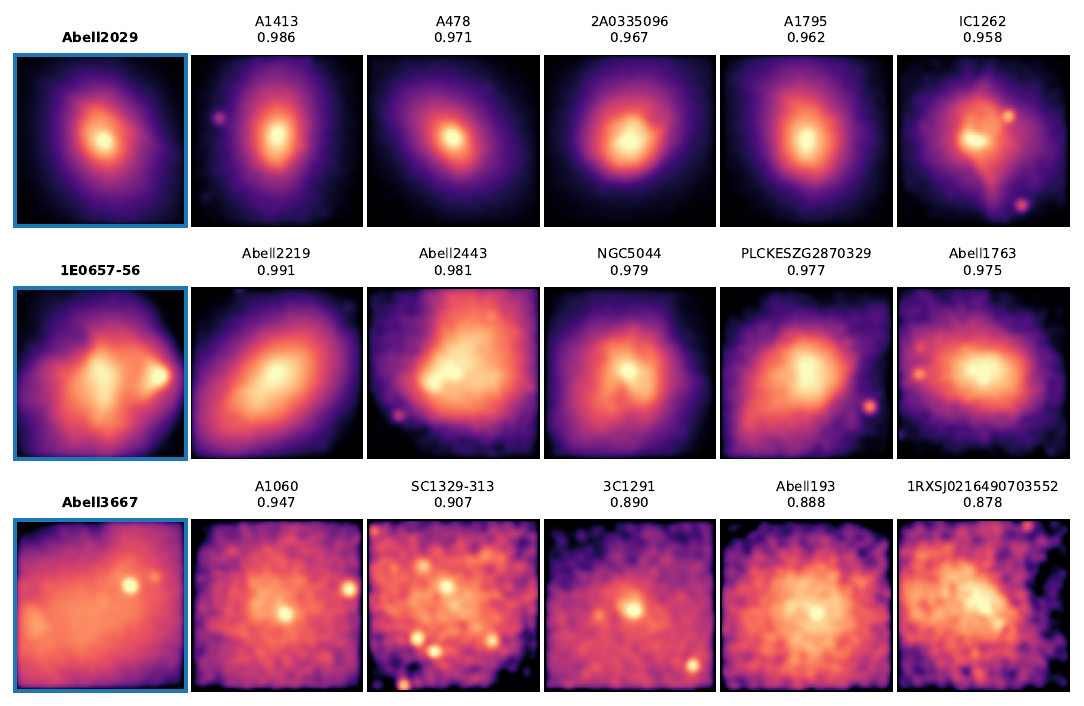}
    \caption{Examples of the similarity search. Each row shows a query object (left, outlined) and its five nearest neighbors in the AstroCLIP embedding space, labeled with their cosine similarity to the query. The images are the encoder inputs: $4.2'\times4.2'$ cutouts of the 0.5--7.0\,keV merged image centered on the X-ray center, resampled, smoothed and stretched as described in Section~\ref{sec:astroclip}.}
    \label{fig:neighbors}
\end{figure*}

The neighbors are not a calibrated physical match, and we assess them only coarsely here. Neighbors are closer in redshift than random pairs of embedded objects: the median $|\Delta \log_{10} z|$ between an object and its neighbors is \neighborDlogZ, compared with \randomDlogZ\ for random pairs. Because galaxies are local and clusters are not, part of this would follow from the two classes alone, but the effect persists among clusters: \neighborDlogZClusters\ for neighboring clusters, compared with \randomDlogZClusters\ for random pairs of clusters. Part of the remaining signal is expected from the fixed angular size of the cutouts, since the apparent extent of the emission decreases with redshift.
Figure~\ref{fig:neighbors} shows three examples. The neighbors of the relaxed cool-core cluster Abell~2029 are clusters and groups with bright, centrally peaked cores, among them the cool-core clusters Abell~478, 2A~0335+096 and Abell~1795, and the neighbors of the merging Bullet Cluster (1E0657-56) include the merging clusters Abell~2219 and PLCK~G287.0+32.9. Abell~3667 illustrates a limitation of the fixed cutout size: the cluster is nearby and its emission fills the $4.2'$ field, so its neighbors are other nearby objects whose emission fills the field, such as Abell~1060 and Abell~3376, rather than clusters with a similar morphology.

To test whether the neighbors reflect physical properties of the gas, we compare them with the core entropies $K_0$ measured by ACCEPT \citep{cavagnolo_intracluster_2009}. We match each ACCEPT cluster to the embedded object whose X-ray center lies closest to the ACCEPT position, within $1'$, keeping one object per ACCEPT cluster so that duplicate observations of the same cluster cannot match each other. This gives \nAcceptMatched\ clusters and \nAcceptPairs\ object--neighbor pairs in which both objects have a $K_0$. We label a cluster a cool core if $K_0<30$\,keV\,cm$^2$; \acceptCoolCorePercent\% of the matched clusters are cool cores. An object and its neighbor share cool-core status in \ccAgreementPercent\% of pairs. To estimate how often this would happen by chance, we shuffle the cool-core labels \nPermutations\ times, shuffling only among clusters in the same redshift quintile because both the cool-core fraction in ACCEPT and the neighbors themselves depend on redshift. The shuffled labels agree in \ccNullPercent\% of pairs on average, and none of the \nPermutations\ shuffles reaches the observed agreement, which exceeds their mean by \ccSignificance\ times their standard deviation. Each object contributes up to \nNeighbors\ pairs, so the pairs are not independent; the shuffles account for this, because they keep the pairs and permute only the labels. The effect is present for both classes: \ccGivenCCPercent\% of the neighbors of cool-core clusters are themselves cool cores, compared with \ccGivenNCCPercent\% of the neighbors of non-cool-core clusters. Although AstroCLIP was never trained on X-ray images, its embeddings therefore capture the central concentration of the emission that distinguishes cool-core clusters.

\subsection{Website}
In order to construct the web-based catalog, we stored all relevant metadata (i.e. ObsIDs, RA/DEC, and redshift) in a \texttt{sqlite3} database. The website is hosted on Cloudflare: a Cloudflare Worker serves the static frontend (a standard HTML/CSS/JS stack with Bootstrap styling) and an API that queries the object database, which is stored in Cloudflare D1. Each object's page shows the postage stamps of its broad, soft and hard images, its ObsIDs, positions and redshift, and its most similar objects in the catalog. The FITS products described in Section~\ref{sec:cleaning} are packaged per object in a single archive served from a Cloudflare R2 bucket, which for most multi-ObsID objects also contains the merged event file. On the online catalog, users can search for individual objects, filter based on search criteria, download the FITS products, access postage stamp images of all objects, and browse the most similar objects in the catalog.

\section{Results} \label{sec:results}
\subsection{Catalog statistics} \label{sec:stats}
% Generated by scripts/paper_stats.py -- do not edit by hand.
\begin{table}[t]
\centering
\caption{The ten deepest objects in the LEMUR catalog: 5 galaxy clusters and 5 galaxies. A dash marks an object without a redshift.}
\label{tab:deepest}
\begin{tabular}{lccc}
\hline
Name & $z$ & $N_{\rm ObsID}$ & Exposure (Ms) \\
\hline
Perseus     & 0.0170 & 121 & 3.32 \\
SgrA        & --- & 67 & 2.48 \\
Abell2146   & 0.2337 & 77 & 2.32 \\
Abell133    & 0.0565 & 33 & 2.29 \\
ComaCluster & 0.0234 & 34 & 1.46 \\
M33         & $-0.0006$ & 30 & 1.43 \\
NGC3115     & 0.0022 & 11 & 1.12 \\
M31         & $-0.0010$ & 135 & 1.02 \\
M101        & 0.0008 & 25 & 0.99 \\
M82         & 0.0007 & 25 & 0.93 \\
\hline
\end{tabular}
\end{table}

\begin{figure*}
    \centering
    \includegraphics[width=\linewidth]{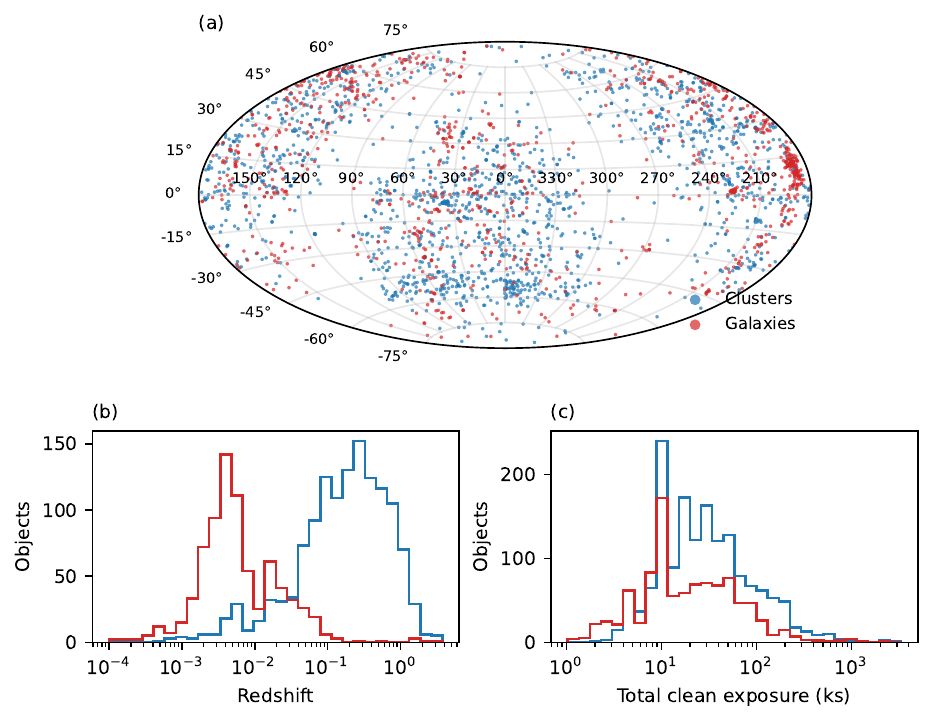}
    \caption{Overview of the catalog. (a) Sky distribution in equatorial coordinates (Aitoff projection). (b) Redshift distribution of the objects with a measured redshift. (c) Total clean exposure per object. Galaxy clusters are shown in blue and galaxies in red in all panels.}
    \label{fig:catalog}
\end{figure*}

The catalog contains \nObjects\ objects, \nClusters\ galaxy clusters and \nGalaxies\ galaxies, built from \nObsIDs\ unique ObsIDs. We classify an object as a cluster when any of its observations was submitted under the CDA science category ``clusters of galaxies'', and as a galaxy otherwise. \nMultiObsID\ objects combine more than one ObsID, with up to \maxObsIDsPerObject\ ObsIDs for a single object. Figure~\ref{fig:catalog}a shows the distribution of the catalog on the sky; the concentration of galaxies near RA\,$\approx190^\circ$, Dec\,$\approx+12^\circ$ is the Virgo cluster region.

A redshift is available for \nWithRedshift\ objects (Figure~\ref{fig:catalog}b). The median redshift is \medianRedshift, and the most distant object is at $z=\maxRedshift$. The galaxies are almost all local, while the clusters extend to $z>1$.

The total clean exposure in the catalog is \totalExposureMs\,Ms (Figure~\ref{fig:catalog}c). The median total exposure per object is \medianExposureKs\,ks, and \nAboveHundredKs\ objects have more than 100\,ks. Table~\ref{tab:deepest} lists the deepest objects, the deepest of which reaches \maxExposureMs\,Ms.

\subsection{Image validation} \label{sec:validation}

\begin{figure*}
    \centering
    \includegraphics[width=0.85\linewidth]{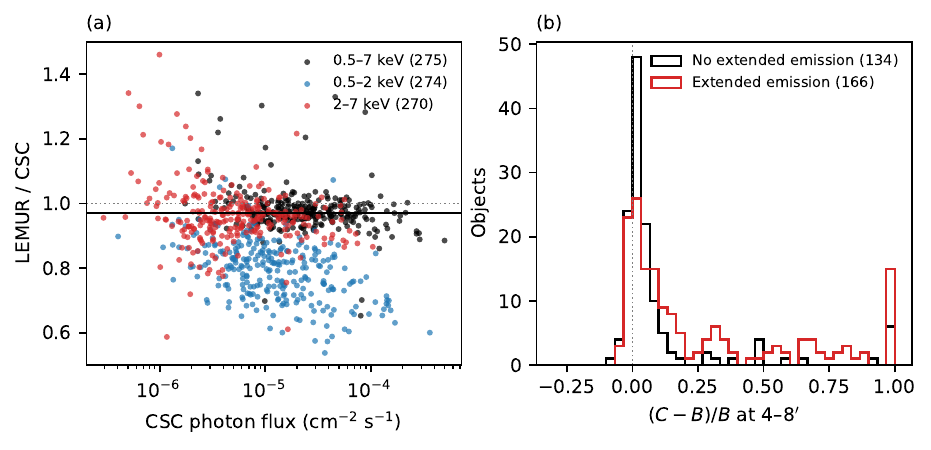}
    \caption{Validation of the merged images. (a) Ratio of the LEMUR to the CSC photon flux of bright, isolated point sources as a function of the CSC flux, in the broad (black), soft (blue) and hard (red) bands; the solid line marks the median broad-band ratio. (b) Residual background $(C-B)/B$ between $4'$ and $8'$ from the target, on chips with a blank-sky background, for objects without (black) and with (red) significant extended emission; values above 1 are shown in the last bin.}
    \label{fig:validation}
\end{figure*}

We test the flux scale and the background subtraction of the merged images against the Chandra Source Catalog Release~2.1 \citep{evans_chandra_2024}, which was built from the same observations with an independent pipeline. We use the objects whose observations all started before 2022, so that the CSC includes them, and that combine at most five ObsIDs. Within $3'$ of each target, we select the CSC point sources detected with a significance above 10 that are not flagged as extended, variable, piled up, confused, saturated or on a readout streak, and that have no other CSC source within $30''$. For each source, we sum the LEMUR image $F$ within $8''$ of the CSC position and subtract the local level measured in a $15$--$30''$ annulus, discarding the sources on bright diffuse emission, for which this local level exceeds 20\% of the aperture flux. Figure~\ref{fig:validation}a compares the photon fluxes of \nValSources\ sources in \nValObjects\ objects with the aperture-corrected CSC photon fluxes in the same bands. In the broad band, the median ratio of the LEMUR to the CSC flux is \valBroadRatio, with 68\% of the sources between \valBroadLow\ and \valBroadHigh; in the hard band it is \valHardRatio\ (\valHardLow--\valHardHigh). A deficit of a few per cent is expected from the fraction of the point-spread function that falls outside the $8''$ aperture.

In the soft band, the LEMUR fluxes are about 20\% lower than the CSC fluxes (median ratio \valSoftRatio, 68\% between \valSoftLow\ and \valSoftHigh), and the deficit is larger for later observations, with a median ratio of \valSoftEarly\ for objects observed before 2005 and \valSoftLate\ for objects observed after 2014, and for softer sources. This is a consequence of computing each exposure map at a single energy. The soft-band exposure map is evaluated at 1.25\,keV, where the effective area is larger than at 0.5--1\,keV, and the difference has grown as contamination has built up on the ACIS optical blocking filters \citep{plucinsky_complicated_2018}; dividing the soft-band counts by this exposure therefore underestimates the photon flux of a typical source. The broad band is much less affected, because the effective area at 2.3\,keV lies between those at the lower and higher energies of the band. Users who need accurate soft-band fluxes should compute them from the counts and background images with a spectral response appropriate to their source.

To test the background subtraction, we measure the residual $(C-B)/B$ between $4'$ and $8'$ from the target in \nBkgObjects\ objects selected at random among those with at most three ObsIDs, all observed before 2022, on chips that have a blank-sky background and more than $20''$ from any CSC source. For the \nBkgQuiet\ objects without significant extended emission (Section~\ref{sec:centers}), the median residual is \bkgQuietMedian\% of the blank-sky level, 68\% of them lie between \bkgQuietLow\% and \bkgQuietHigh\%, and \bkgQuietWithinTen\% are within 10\% (Figure~\ref{fig:validation}b). A residual of a few per cent is expected, because the blank-sky fields do not match the Galactic foreground and cosmic X-ray background of each target exactly. Objects with extended emission have larger residuals because their own emission extends beyond $4'$, and the largest residuals in both groups belong to galaxies and pointings inside bright nearby clusters such as Coma, Perseus and Virgo, whose emission fills the field. In \pctNoBkgModel\% of the objects, at least one chip has no blank-sky background and is therefore not background subtracted (Section~\ref{sec:cleaning}).

\IfFileExists{showcase_macros.tex}{% Generated by scripts/showcase_figures.py -- do not edit by hand.
\newcommand{\ngcFourSixThreeSixObsIDs}{323, 324, 3926, 4415}
\newcommand{\ngcFourSixThreeSixNObsIDs}{4}
\newcommand{\ngcFourSixThreeSixExposureKs}{209}
\newcommand{\ngcFourSixThreeSixCenterRA}{12:42:49.51}
\newcommand{\ngcFourSixThreeSixCenterDec}{+02:41:03.0}
\newcommand{\ngcFourSixThreeSixRedshift}{0.00307}
\newcommand{\ngcFourSixThreeSixMissingObsIDs}{none}
\newcommand{\bulletObsIDs}{554, 3184, 4984, 4985, 4986, 5355, 5356, 5357, 5358, 5361}
\newcommand{\bulletNObsIDs}{10}
\newcommand{\bulletExposureKs}{533}
\newcommand{\bulletCenterRA}{06:58:30.53}
\newcommand{\bulletCenterDec}{$-$55:56:25.3}
\newcommand{\bulletRedshift}{0.2956}
\newcommand{\bulletMissingObsIDs}{none}
}{%
  \newcommand{\ngcFourSixThreeSixObsIDs}{TBD}%
  \newcommand{\ngcFourSixThreeSixNObsIDs}{TBD}%
  \newcommand{\ngcFourSixThreeSixExposureKs}{TBD}%
  \newcommand{\ngcFourSixThreeSixCenterRA}{TBD}%
  \newcommand{\ngcFourSixThreeSixCenterDec}{TBD}%
  \newcommand{\ngcFourSixThreeSixRedshift}{TBD}%
  \newcommand{\ngcFourSixThreeSixMissingObsIDs}{TBD}%
  \newcommand{\bulletObsIDs}{TBD}%
  \newcommand{\bulletNObsIDs}{TBD}%
  \newcommand{\bulletExposureKs}{TBD}%
  \newcommand{\bulletCenterRA}{TBD}%
  \newcommand{\bulletCenterDec}{TBD}%
  \newcommand{\bulletRedshift}{TBD}%
  \newcommand{\bulletMissingObsIDs}{TBD}%
}

\subsection{Showcase reductions} \label{sec:showcase}

To illustrate what the catalog provides, we compare the reductions of two well-studied objects, the elliptical galaxy NGC\,4636 and the merging Bullet Cluster (1E0657-56), with the deepest published Chandra analyses of each. For both objects, the catalog merge includes every public ACIS observation in the archive: a search of the Chandra observation catalog within $40'$ of each target returns no observed ACIS ObsIDs beyond those we merge.
% Archive check: CDA ocat (cda.harvard.edu/srservices, 54,884 rows), cone
% searches of 15' and 15-40' around each target. Four more Bullet ObsIDs
% (21719, 22719, 23025, 23028; ACIS-I, 50 ks each) are listed but "canceled".
% No HRC or grating observations of either target.

\subsubsection{NGC\,4636}

\begin{figure}
    \centering
    \includegraphics[width=0.95\linewidth]{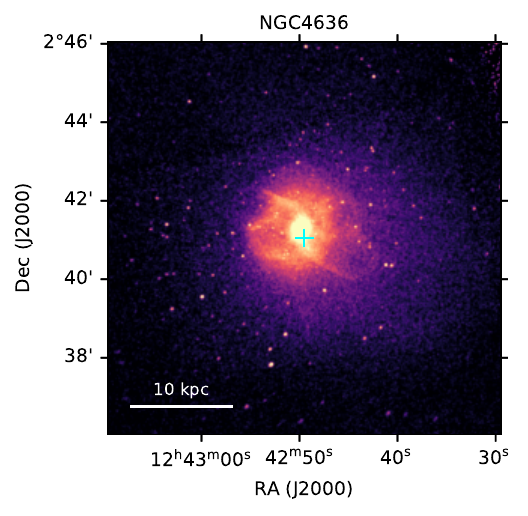}
    \caption{Merged, background-subtracted, exposure-corrected 0.5--7.0\,keV Chandra image of the elliptical galaxy NGC\,4636 ($z=\ngcFourSixThreeSixRedshift$), combining ObsIDs \ngcFourSixThreeSixObsIDs\ (\ngcFourSixThreeSixExposureKs\,ks of clean exposure). The image is smoothed with a Gaussian of $\sigma=1''$ and shown with an $\mathrm{arcsinh}$ stretch ($a=0.08$). The cross marks the X-ray center (\ngcFourSixThreeSixCenterRA, \ngcFourSixThreeSixCenterDec; Section~\ref{sec:centers}). The scale bar assumes the Planck 2018 cosmology \citepalias{planck_collaboration_planck_2020}.}
    \label{fig:ngc4636}
\end{figure}

NGC\,4636 is a nearby X-ray-bright elliptical galaxy in the outskirts of the Virgo cluster, whose hot gas shows arm-like features attributed to past nuclear outbursts \citep{jones_chandra_2002}.
% Checked 2026-11-18 against jones_chandra_2002 (astro-ph/0108114): Sec. 1 "NGC 4636
% lies in the outskirts of the Virgo cluster, 10 deg or 2.6 Mpc on the sky to the
% south of M87"; abstract: the arm-like structures "are produced by nuclear outburst
% driven shocks".
Chandra observed it four times: once with ACIS-S in 2000 (ObsID 323, 52\,ks) and three times with ACIS-I (ObsIDs 324, 3926 and 4415, of 8, 75 and 74\,ks), for 210\,ks in total before cleaning.
% Raw archive exposures (ocat EXP_TIME): 323 52.38 ks (ACIS-S, FAINT,
% 2000-01-26); 324 8.38 ks (ACIS-I, VFAINT, 1999-12-04); 3926 74.70 ks and
% 4415 74.36 ks (ACIS-I, 2003-02-14/15); total 209.8 ks.
The first Chandra studies used only the 1999--2000 data: \citet{jones_chandra_2002} analyzed ObsIDs 323 and 324 with 41 and 6\,ks of good time, and \citet{osullivan_active_2005} used 38.9\,ks of ObsID 323 after cleaning.
% jones_chandra_2002 (2002ApJ...567L.115J) Sec. 2: good times 41286 s (323) and
% 5989 s (324). It quotes 11 ks for 324 raw; the archive gives 8.38 ks.
% osullivan_active_2005 (2005ApJ...624L..77O) Sec. 2: "~38.9 ks for ACIS-S".
The deepest imaging studies combine the 2000 ACIS-S observation with the two 2003 ACIS-I observations: \citet{baldi_unusual_2009} reach about 186\,ks of good time (37\,ks ACIS-S and 149\,ks ACIS-I) and \citet{posson-brown_chandra_2009} 193\,ks.
% baldi_unusual_2009 (2009ApJ...707.1034B) abstract "~200ks in total"; Sec. 2:
% usable ~37 ks (ACIS-S) and ~149 ks (ACIS-I). Its Sec. 2 says "twice by
% ACIS-S", but the archive has one ACIS-S ObsID (323).
% posson-brown_chandra_2009 (2009ApJ...695.1094P) Table 1: lc_clean exposures 44,450,
% 74,709 and 74,190 s (sum 193.3 ks); the 5 ks 1999 ACIS-I data were not used.
The Chandra Galaxy Atlas \citep{kim_chandra_2019} is the only work we found that merges the same four ObsIDs as our catalog, with 191.1\,ks after flare filtering.
% kim_chandra_2019 (2019ApJS..241...36K) Table 1 col. 13 (N4636, merge 90401):
% 191.1 ks = 42.4 + 5.5 + 71.4 + 71.8 ks, summed flare-filtered LIVETIME of
% the chip containing the galaxy center.
Our merged image (Figure~\ref{fig:ngc4636}) combines all \ngcFourSixThreeSixNObsIDs\ ObsIDs, the same data set as the Chandra Galaxy Atlas, with a total exposure of \ngcFourSixThreeSixExposureKs\,ks. This is 18\,ks more than \citet{kim_chandra_2019} retain after flare filtering, because our conservative light-curve filtering (Section~\ref{sec:cleaning}) removed only 1.2\,ks (0.6\%) from these four observations and none from ObsID 323, in which \citet{jones_chandra_2002} excluded 11\,ks of background flares. In exposure, the catalog image therefore matches the deepest published reductions; users who need flare-free data, for example for spectroscopy of the faint outer halo, should apply additional flare filtering to ObsID 323.
% Per-ObsID level-2 EXPOSURE -> our repro evt2 EXPOSURE (2026 reduction):
% 323 52.4 -> 52.4; 324 8.4 -> 8.4; 3926 74.7 -> 73.9; 4415 74.4 -> 74.0 ks;
% total 209.8 -> 208.6 ks. 191.1 ks from kim_chandra_2019 Table 1 col. 13;
% jones_chandra_2002 good time for 323: 41.3 of 52.4 ks.

\subsubsection{The Bullet Cluster (1E0657-56)}

\begin{figure}
    \centering
    \includegraphics[width=0.95\linewidth]{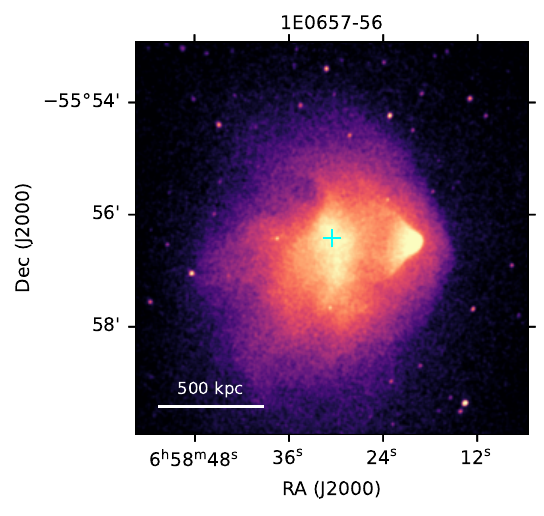}
    \caption{Merged, background-subtracted, exposure-corrected 0.5--7.0\,keV Chandra image of the Bullet Cluster, 1E0657-56 ($z=\bulletRedshift$), combining ObsIDs \bulletObsIDs\ (\bulletExposureKs\,ks of clean exposure). The image is smoothed with a Gaussian of $\sigma=1''$ and shown with an $\mathrm{arcsinh}$ stretch ($a=0.08$). The cross marks the X-ray center (\bulletCenterRA, \bulletCenterDec; Section~\ref{sec:centers}), which falls on the main cluster rather than the bullet subcluster. The scale bar assumes the Planck 2018 cosmology \citepalias{planck_collaboration_planck_2020}.}
    \label{fig:bullet}
\end{figure}

The Bullet Cluster is a merging cluster at $z=\bulletRedshift$ in which a dense subcluster has passed through the main cluster, driving a bow shock ahead of its cool core \citep{markevitch_textbook_2002}. It was first observed by Chandra in October 2000 (ObsID 554) and again in 2002 (ObsID 3184), followed by eight ACIS-I observations in August 2004, for a total of ten ACIS-I ObsIDs and 577\,ks before cleaning.
% Raw archive exposures (ocat EXP_TIME, all ACIS-I): 554 25.79 (FAINT,
% 2000-10-16); 3184 87.47 (VFAINT, 2002-07-12); 5355 27.41; 5356 97.20;
% 5357 79.06; 5358 31.96; 5361 82.62; 4984 76.12; 4985 27.49; 4986 41.49
% (all VFAINT, 2004-08-10..25); total 576.6 ks.
The discovery of the shock used a single 24.3\,ks observation \citep{markevitch_textbook_2002}, and a 70\,ks observation in 2002 raised the depth to about 94\,ks \citep{markevitch_direct_2004}.
% markevitch_textbook_2002 (2002ApJ...567L..27M) Sec. 2: "useful exposure of 24.3 ks";
% Sec. 1 dates it to October 2000, i.e. ObsID 554 (matched by date; the
% paper does not name ObsIDs).
% markevitch_direct_2004 (2004ApJ...606..819M) Sec. 1: 24 ks (Oct 2000) and 70 ks
% (July 2002), i.e. 554 and 3184 by date; raw or clean not stated.
Together with the 2004 campaign, these data formed the well-known 500\,ks image used to demonstrate the offset between the gas and the lensing mass \citep{markevitch_2006,clowe_direct_2006}.
% markevitch_2006 (2006ESASP.604..723M, confirmed on ADS) abstract and Sec. 1:
% "500 ks"; clowe_direct_2006 (2006ApJ...648L.109C) Fig. 1 caption: "a 500 ks
% Chandra image". Neither lists ObsIDs or says whether 500 ks is cleaned.
Among later studies that state their depth, \citet{shimwell_another_2015} used 522\,ks of clean exposure, and \citet{liu_method_2015} merged nine of the ten ObsIDs, excluding the FAINT-mode ObsID 554, for about 550\,ks after data reduction.
% shimwell_another_2015 (2015MNRAS.449.1486S) Sec. 2: "total clean exposure of 522 ks".
% liu_method_2015 (2015ApJ...809...27L) Sec. 4.1 and Table 2: 3184, 5355-5358, 5361,
% 4984-4986; "The total exposure time of the merged data is ~550.4 ks".
Our merged image (Figure~\ref{fig:bullet}) includes all \bulletNObsIDs\ ObsIDs, including the FAINT-mode ObsID 554 that \citet{liu_method_2015} excluded, with a total exposure of \bulletExposureKs\,ks. Our light-curve filtering removed 44\,ks (7.6\%) of the 577\,ks archive exposure, 32\,ks of it from ObsID 5361, whereas \citet{liu_method_2015} retained nearly all of the exposure of their nine ObsIDs (550 of 551\,ks). The catalog image therefore reaches a depth comparable to the deepest published analyses, above the 522\,ks of \citet{shimwell_another_2015} and about 3\% below the 550\,ks of \citet{liu_method_2015}.
% Per-ObsID level-2 EXPOSURE -> our repro evt2 EXPOSURE (2026 reduction), ks:
% 554 25.8->25.8; 3184 87.5->83.1; 4984 76.1->75.7; 4985 27.5->23.9;
% 4986 41.5->41.5; 5355 27.4->24.6; 5356 97.2->96.8; 5357 79.1->78.9;
% 5358 32.0->32.0; 5361 82.6->50.3; total 576.6 -> 532.6 ks.
% Liu's nine ObsIDs total 550.8 ks in the archive; they quote ~550.4 ks.

\section{Conclusions} \label{sec:conclusion}

We have presented LEMUR, a publicly accessible archive of uniformly reduced, deep \textit{Chandra} images of \nObjects\ galaxies and galaxy clusters, comprising \nObsIDs\ ACIS observations and \totalExposureMs\,Ms of exposure. By combining all publicly available imaging observations of each target, apart from calibration pointings and offset fields, we provide background-subtracted, exposure-corrected images at the native ACIS pixel scale of 0.492''. Our primary goal is to make the wealth of archival \textit{Chandra} observations accessible to researchers without requiring specialized expertise in X-ray data reduction. Through the LEMUR web interface, users can browse the catalog, inspect images, and download the corresponding FITS products. The catalog also provides both the original target positions and measured X-ray centers, facilitating cross-matching with multiwavelength observations.

An additional feature of LEMUR is its morphological similarity search, which uses embeddings from the AstroCLIP foundation model to identify objects with similar X-ray emission structures. Despite having been trained on optical observations, AstroCLIP identifies physically meaningful similarities between galaxy clusters: nearest neighbors share cool-core classifications in \ccAgreementPercent\% of matched pairs, compared with \ccNullPercent\% expected from redshift-stratified randomization. This demonstrates the potential of pretrained astronomical foundation models for organizing archival observations by morphology, even when applied outside their original training domain. More broadly, the uniformly processed image collection provides a starting point for machine-learning applications requiring large samples of X-ray images.

Our comparisons with published reductions of NGC 4636 and the Bullet Cluster demonstrate that LEMUR achieves imaging depths comparable to those of dedicated studies while applying a consistent reduction procedure across the archive. This uniformity necessarily involves trade-offs. Whereas bespoke analyses can optimize background filtering, calibration, and image processing for individual targets, LEMUR prioritizes reproducibility, completeness, and the preservation of faint diffuse emission. In particular, our conservative flare filtering can retain intervals of elevated background that would be excluded in analyses optimized for faint-source spectroscopy. The resulting images are therefore intended as accessible, science-ready imaging products and starting points for further analysis, rather than replacements for specialized reductions tailored to individual scientific objectives.

Several limitations should be considered when using the catalog. Grouping observations by target names and sky positions can leave offset observations separate or occasionally combine distinct targets, while the completeness of the catalog is limited by the science categories assigned to observations in the Chandra Data Archive. Additionally, the images of the three widest mosaics (Centaurus, M31 and Perseus) include only the observations within $30'$ of the core, and our automated X-ray center measurements may be less reliable for faint, disturbed, or complex systems. Objects without significant extended emission, \nCentersFallback\ of the \nObjects\ and most of them galaxies, are assigned their target position, and the X-ray center of a galaxy can fall on another source within $1'$ of the target.
 The blank-sky background leaves residuals of a few per cent of the background level, some chips have no blank-sky background, and the soft-band images underestimate photon fluxes by about 20\% because their exposure maps are computed at a single energy (Section~\ref{sec:validation}). Finally, the morphological similarity search uses a model trained on optical images and fixed-angular-size cutouts, introducing sensitivity to apparent source extent and redshift. Consequently, similarity rankings should be interpreted as morphological associations rather than calibrated measures of physical similarity.

Future developments will focus on extending both the scientific products and the functionality of the archive. Planned additions include spatially resolved spectral products, physical-scale image cutouts, and embeddings trained specifically on X-ray observations. We will also compare the AstroCLIP embeddings of the observed objects with those of simulated galaxies and clusters. We also intend to automate the incorporation of newly available \textit{Chandra} observations, provide more detailed provenance information for individual reductions, and establish cross-links with complementary X-ray surveys such as \textit{XMM-Newton} and \textit{eROSITA}. Together, these developments will expand LEMUR's utility as an accessible resource for multiwavelength studies and data-driven investigations of galaxies and galaxy clusters.

% Force any pending floats (e.g., the two showcase figures) to be placed
% before the back-matter, to avoid an imbalanced/empty left column.
\clearpage

%% Please use the acknowledgment and contribution environments. This will
%% be anonymized when the "anonymous" style option is used.
\begin{acknowledgments}
This research has made use of data obtained from the Chandra Data Archive provided by the Chandra X-ray Center (CXC). This research has made use of the NASA/IPAC Extragalactic Database (NED), which is funded by the National Aeronautics and Space Administration and operated by the California Institute of Technology. This research has made use of the SIMBAD database \citep{wenger_simbad_2000} and the Sesame name resolver, operated at CDS, Strasbourg, France.

\end{acknowledgments}

\section*{AI Disclosure}
Codex was used extensively to build the web application; each line of its frontend and backend was reviewed by a team member with web-development expertise. Claude Code was used throughout the project: to write and debug parts of the reduction pipeline and the catalog tooling (including the X-ray center finder, the reconciliation of the catalog with the archive and the redshift look-ups), to write the scripts that compute the catalog statistics, figures and image validation, to help develop the website, and to draft and revise parts of the manuscript. All code and text produced with these tools were reviewed and edited by the authors.

\section*{Data Availability}
The images, counts images, background models, exposure maps and postage stamps of every object are available from the LEMUR web archive at \url{https://lemur-archive.lemur-xray.workers.dev/}. The catalog tables of this data release (DR1), with the target positions, X-ray centers, redshifts, ObsIDs and clean exposures of all objects, their most similar objects and their AstroCLIP embeddings, are archived on Zenodo at \href{https://doi.org/10.5281/zenodo.24351915}{doi:10.5281/zenodo.24351915}. The code that builds the catalog and the website is available at \url{https://github.com/crhea93/Lemur} and archived at \href{https://doi.org/10.5281/zenodo.18728228}{doi:10.5281/zenodo.18728228}.

\section*{Software}
Python \citep{van_rossum_python_2009}, Astropy \citepalias{collaboration_astropy_2022}, NumPy \citep{harris_array_2020}, SciPy \citep{virtanen_scipy_2020}, pandas \citep{mckinney_data_2010,pandas_development_team_2026}, Matplotlib \citep{hunter_matplotlib_2007}, PyTorch \citep{paszke_pytorch_2019}, FastAPI \citep{ramirez_fastapi_2018}, CIAO \citep{fruscione_ciao_2026}

\bibliographystyle{aasjournalv7}
\bibliography{lemur}

@inproceedings{garmire_advanced_2003,
	booktitle = {X-{Ray} and {Gamma}-{Ray} {Telescopes} and {Instruments} for {Astronomy}},
	series = {Society of {Photo}-{Optical} {Instrumentation} {Engineers} ({SPIE}) {Conference} {Series}},
	editor = {Truemper, Joachim E. and Tananbaum, Harvey D.},
	title = {Advanced {CCD} imaging spectrometer ({ACIS}) instrument on the {Chandra} {X}-ray {Observatory}},
	volume = {4851},
	url = {https://ui.adsabs.harvard.edu/abs/2003SPIE.4851...28G},
	doi = {10.1117/12.461599},
	urldate = {2026-03-10},
	author = {Garmire, Gordon P. and Bautz, Mark W. and Ford, Peter G. and Nousek, John A. and Ricker, Jr., George R.},
	month = mar,
	year = {2003},
	note = {ADS Bibcode: 2003SPIE.4851...28G},
	pages = {28--44},
}

@inproceedings{white_high_2009,
	booktitle = {astro2010: {The} {Astronomy} and {Astrophysics} {Decadal} {Survey}, {Position} {Papers}},
	title = {The {High} {Impact} of {Astronomical} {Data} {Archives}},
	volume = {2010},
	url = {https://ui.adsabs.harvard.edu/abs/2009astro2010P..64W},
	urldate = {2026-03-10},
	author = {White, Richard L. and Accomazzi, Alberto and Berriman, G. Bruce and Fabbiano, Giuseppina and Madore, Barry F. and Mazzarella, Joseph M. and Rots, Arnold and Smale, Alan P. and Storrie-Lombardi, Lisa and Winkelman, Sherry},
	month = jan,
	year = {2009},
	note = {ADS Bibcode: 2009astro2010P..64W},
	pages = {P64},
}

@article{pandya_comprehensive_2016,
	title = {A Comprehensive Archival {Chandra} Search for {X}-ray Emission from Ultracompact Dwarf Galaxies},
	volume = {819},
	issn = {0004-637X},
	url = {https://doi.org/10.3847/0004-637X/819/2/162},
	doi = {10.3847/0004-637X/819/2/162},
	language = {en},
	number = {2},
	urldate = {2026-03-10},
	journal = {ApJ},
	publisher = {The American Astronomical Society},
	author = {Pandya, Viraj and Mulchaey, John and Greene, Jenny E.},
	month = mar,
	year = {2016},
	pages = {162},
}

@article{hodges-kluck_chandra_2010,
	title = {The \textit{{Chandra}} View of Nearby {X}-shaped Radio Galaxies},
	volume = {710},
	issn = {0004-637X, 1538-4357},
	url = {https://iopscience.iop.org/article/10.1088/0004-637X/710/2/1205},
	doi = {10.1088/0004-637X/710/2/1205},
	language = {en},
	number = {2},
	urldate = {2026-03-10},
	journal = {ApJ},
	author = {Hodges-Kluck, Edmund J. and Reynolds, Christopher S. and Cheung, Chi C. and Miller, M. Coleman},
	month = feb,
	year = {2010},
	pages = {1205--1227},
}

@article{ott_chandra_2005,
	title = {A {Chandra} {X}-ray survey of nearby dwarf starburst galaxies — {I}. {Data} reduction and results},
	volume = {358},
	issn = {0035-8711},
	url = {https://doi.org/10.1111/j.1365-2966.2005.08862.x},
	doi = {10.1111/j.1365-2966.2005.08862.x},
	number = {4},
	urldate = {2026-03-10},
	journal = {Mon Not R Astron Soc},
	author = {Ott, Jürgen and Walter, Fabian and Brinks, Elias},
	month = apr,
	year = {2005},
	pages = {1423--1452},
}

@article{owers_high_2009,
	title = {A High Fidelity Sample of Cold Front Clusters from the {Chandra} Archive},
	volume = {704},
	issn = {0004-637X},
	url = {https://doi.org/10.1088/0004-637X/704/2/1349},
	doi = {10.1088/0004-637X/704/2/1349},
	language = {en},
	number = {2},
	urldate = {2026-03-10},
	journal = {ApJ},
	publisher = {The American Astronomical Society},
	author = {Owers, Matt S. and Nulsen, Paul E. J. and Couch, Warrick J. and Markevitch, Maxim},
	month = oct,
	year = {2009},
	pages = {1349},
}

@article{cavagnolo_intracluster_2009,
	title = {Intracluster {Medium} {Entropy} {Profiles} for a {Chandra} {Archival} {Sample} of {Galaxy} {Clusters}},
	volume = {182},
	issn = {0067-0049},
	url = {https://ui.adsabs.harvard.edu/abs/2009ApJS..182...12C},
	doi = {10.1088/0067-0049/182/1/12},
	urldate = {2026-03-09},
	journal = {The Astrophysical Journal Supplement Series},
	publisher = {IOP},
	author = {Cavagnolo, Kenneth W. and Donahue, Megan and Voit, G. Mark and Sun, Ming},
	month = may,
	year = {2009},
	note = {ADS Bibcode: 2009ApJS..182...12C},
	pages = {12--32},
}

@misc{paszke_pytorch_2019,
	title = {{PyTorch}: {An} {Imperative} {Style}, {High}-{Performance} {Deep} {Learning} {Library}},
	shorttitle = {{PyTorch}},
	url = {https://ui.adsabs.harvard.edu/abs/2019arXiv191201703P},
	doi = {10.48550/arXiv.1912.01703},
	urldate = {2025-12-08},
	publisher = {arXiv},
	author = {Paszke, Adam and Gross, Sam and Massa, Francisco and Lerer, Adam and Bradbury, James and Chanan, Gregory and Killeen, Trevor and Lin, Zeming and Gimelshein, Natalia and Antiga, Luca and Desmaison, Alban and Köpf, Andreas and Yang, Edward and DeVito, Zach and Raison, Martin and Tejani, Alykhan and Chilamkurthy, Sasank and Steiner, Benoit and Fang, Lu and Bai, Junjie and Chintala, Soumith},
	month = dec,
	year = {2019},
	note = {ADS Bibcode: 2019arXiv191201703P},
}

@article{hunter_matplotlib_2007,
	title = {Matplotlib: {A} {2D} {Graphics} {Environment}},
	volume = {9},
	issn = {1558-366X},
	shorttitle = {Matplotlib},
	doi = {10.1109/MCSE.2007.55},
	number = {3},
	journal = {Computing in Science Engineering},
	author = {Hunter, John D.},
	year = {2007},
	note = {Conference Name: Computing in Science Engineering},
	pages = {90--95},
}

@article{ebeling_detecting_1993,
	title = {Detecting structure in two dimensions combining {Voronoi} tessellation and percolation},
	volume = {47},
	issn = {1063-651X},
	url = {https://ui.adsabs.harvard.edu/abs/1993PhRvE..47..704E},
	doi = {10.1103/PhysRevE.47.704},
	urldate = {2026-10-14},
	journal = {Physical Review E},
	publisher = {APS},
	author = {Ebeling, H. and Wiedenmann, G.},
	month = jan,
	year = {1993},
	note = {ADS Bibcode: 1993PhRvE..47..704E},
	pages = {704--710},
}

@article{merloni_srgerosita_2024,
	title = {The {SRG}/{eROSITA} all-sky survey. {First} {X}-ray catalogues and data release of the western {Galactic} hemisphere},
	volume = {682},
	issn = {0004-6361},
	url = {https://ui.adsabs.harvard.edu/abs/2024A&A...682A..34M},
	doi = {10.1051/0004-6361/202347165},
	urldate = {2026-09-29},
	journal = {Astronomy and Astrophysics},
	publisher = {EDP},
	author = {Merloni, A. and Lamer, G. and Liu, T. and Ramos-Ceja, M. E. and Brunner, H. and Bulbul, E. and Dennerl, K. and Doroshenko, V. and Freyberg, M. J. and Friedrich, S. and Gatuzz, E. and Georgakakis, A. and Haberl, F. and Igo, Z. and Kreykenbohm, I. and Liu, A. and Maitra, C. and Malyali, A. and Mayer, M. G. F. and Nandra, K. and Predehl, P. and Robrade, J. and Salvato, M. and Sanders, J. S. and Stewart, I. and Tubín-Arenas, D. and Weber, P. and Wilms, J. and Arcodia, R. and Artis, E. and Aschersleben, J. and Avakyan, A. and Aydar, C. and Bahar, Y. E. and Balzer, F. and Becker, W. and Berger, K. and Boller, T. and Bornemann, W. and Brüggen, M. and Brusa, M. and Buchner, J. and Burwitz, V. and Camilloni, F. and Clerc, N. and Comparat, J. and Coutinho, D. and Czesla, S. and Dannhauer, S. M. and Dauner, L. and Dauser, T. and Dietl, J. and Dolag, K. and Dwelly, T. and Egg, K. and Ehl, E. and Freund, S. and Friedrich, P. and Gaida, R. and Garrel, C. and Ghirardini, V. and Gokus, A. and Grünwald, G. and Grandis, S. and Grotova, I. and Gruen, D. and Gueguen, A. and Hämmerich, S. and Hamaus, N. and Hasinger, G. and Haubner, K. and Homan, D. and Ider Chitham, J. and Joseph, W. M. and Joyce, A. and König, O. and Kaltenbrunner, D. M. and Khokhriakova, A. and Kink, W. and Kirsch, C. and Kluge, M. and Knies, J. and Krippendorf, S. and Krumpe, M. and Kurpas, J. and Li, P. and Liu, Z. and Locatelli, N. and Lorenz, M. and Müller, S. and Magaudda, E. and Mannes, C. and McCall, H. and Meidinger, N. and Michailidis, M. and Migkas, K. and Muñoz-Giraldo, D. and Musiimenta, B. and Nguyen-Dang, N. T. and Ni, Q. and Olechowska, A. and Ota, N. and Pacaud, F. and Pasini, T. and Perinati, E. and Pires, A. M. and Pommranz, C. and Ponti, G. and Poppenhaeger, K. and Pühlhofer, G. and Rau, A. and Reh, M. and Reiprich, T. H. and Roster, W. and Saeedi, S. and Santangelo, A. and Sasaki, M. and Schmitt, J. and Schneider, P. C. and Schrabback, T. and Schuster, N. and Schwope, A. and Seppi, R. and Serim, M. M. and Shreeram, S. and Sokolova-Lapa, E. and Starck, H. and Stelzer, B. and Stierhof, J. and Suleimanov, V. and Tenzer, C. and Traulsen, I. and Trümper, J. and Tsuge, K. and Urrutia, T. and Veronica, A. and Waddell, S. G. H. and Willer, R. and Wolf, J. and Yeung, M. C. H. and Zainab, A. and Zangrandi, F. and Zhang, X. and Zhang, Y. and Zheng, X.},
	month = feb,
	year = {2024},
	note = {ADS Bibcode: 2024A\&A...682A..34M},
	pages = {A34},
}

@article{mehrtens_xmm_2012,
	title = {The {XMM} {Cluster} {Survey}: optical analysis methodology and the first data release: {The} {XMM} {Cluster} {Survey} first data release},
	volume = {423},
	issn = {00358711},
	shorttitle = {The {XMM} {Cluster} {Survey}},
	url = {https://academic.oup.com/mnras/article-lookup/doi/10.1111/j.1365-2966.2012.20931.x},
	doi = {10.1111/j.1365-2966.2012.20931.x},
	language = {en},
	number = {2},
	urldate = {2026-09-29},
	journal = {Monthly Notices of the Royal Astronomical Society},
	author = {Mehrtens, Nicola and Romer, A. Kathy and Hilton, Matt and Lloyd-Davies, E. J. and Miller, Christopher J. and Stanford, S. A. and Hosmer, Mark and Hoyle, Ben and Collins, Chris A. and Liddle, Andrew R. and Viana, Pedro T. P. and Nichol, Robert C. and Stott, John P. and Dubois, E. Naomi and Kay, Scott T. and Sahlén, Martin and Young, Owain and Short, C. J. and Christodoulou, L. and Watson, William A. and Davidson, Michael and Harrison, Craig D. and Baruah, Leon and Smith, Mathew and Burke, Claire and Mayers, Julian A. and Deadman, Paul-James and Rooney, Philip J. and Edmondson, Edward M. and West, Michael and Campbell, Heather C. and Edge, Alastair C. and Mann, Robert G. and Sabirli, Kivanc and Wake, David and Benoist, Christophe and Da Costa, Luiz and Maia, Marcio A. G. and Ogando, Ricardo},
	month = jun,
	year = {2012},
	pages = {1024--1052},
}

@article{romer_serendipitous_2001,
	title = {A {Serendipitous} {Galaxy} {Cluster} {Survey} with \textit{{XMM}} : {Expected} {Catalog} {Properties} and {Scientific} {Applications}},
	volume = {547},
	issn = {0004-637X, 1538-4357},
	shorttitle = {A {Serendipitous} {Galaxy} {Cluster} {Survey} with \textit{{XMM}}},
	url = {https://iopscience.iop.org/article/10.1086/318382},
	doi = {10.1086/318382},
	language = {en},
	number = {2},
	urldate = {2026-09-29},
	journal = {ApJ},
	author = {Romer, A. Kathy and Viana, Pedro T. P. and Liddle, Andrew R. and Mann, Robert G.},
	month = feb,
	year = {2001},
	pages = {594--608},
}

@article{bulbul_srgerosita_2024,
	title = {The {SRG}/{eROSITA} {All}-{Sky} {Survey}. {The} first catalog of galaxy clusters and groups in the {Western} {Galactic} {Hemisphere}},
	volume = {685},
	issn = {0004-6361},
	url = {https://ui.adsabs.harvard.edu/abs/2024A&A...685A.106B},
	doi = {10.1051/0004-6361/202348264},
	urldate = {2026-09-29},
	journal = {Astronomy and Astrophysics},
	publisher = {EDP},
	author = {Bulbul, E. and Liu, A. and Kluge, M. and Zhang, X. and Sanders, J. S. and Bahar, Y. E. and Ghirardini, V. and Artis, E. and Seppi, R. and Garrel, C. and Ramos-Ceja, M. E. and Comparat, J. and Balzer, F. and Böckmann, K. and Brüggen, M. and Clerc, N. and Dennerl, K. and Dolag, K. and Freyberg, M. and Grandis, S. and Gruen, D. and Kleinebreil, F. and Krippendorf, S. and Lamer, G. and Merloni, A. and Migkas, K. and Nandra, K. and Pacaud, F. and Predehl, P. and Reiprich, T. H. and Schrabback, T. and Veronica, A. and Weller, J. and Zelmer, S.},
	month = may,
	year = {2024},
	note = {ADS Bibcode: 2024A\&A...685A.106B},
	pages = {A106},
}

@article{evans_chandra_2024,
	title = {The {Chandra} {Source} {Catalog} {Release} 2 {Series}},
	volume = {274},
	issn = {0067-0049},
	url = {https://ui.adsabs.harvard.edu/abs/2024ApJS..274...22E},
	doi = {10.3847/1538-4365/ad6319},
	urldate = {2026-09-29},
	journal = {The Astrophysical Journal Supplement Series},
	publisher = {IOP},
	author = {Evans, Ian N. and Evans, Janet D. and Martínez-Galarza, J. Rafael and Miller, Joseph B. and Primini, Francis A. and Azadi, Mojegan and Burke, Douglas J. and Civano, Francesca M. and D'Abrusco, Raffaele and Fabbiano, Giuseppina and Graessle, Dale E. and Grier, John D. and Houck, John C. and Lauer, Jennifer and McCollough, Michael L. and Nowak, Michael A. and Plummer, David A. and Rots, Arnold H. and Siemiginowska, Aneta and Tibbetts, Michael S.},
	month = oct,
	year = {2024},
	note = {ADS Bibcode: 2024ApJS..274...22E},
	pages = {22},
}

@article{evans_chandra_2010,
	title = {The {Chandra} {Source} {Catalog}},
	volume = {189},
	issn = {0067-0049},
	url = {https://ui.adsabs.harvard.edu/abs/2010ApJS..189...37E},
	doi = {10.1088/0067-0049/189/1/37},
	urldate = {2026-09-29},
	journal = {The Astrophysical Journal Supplement Series},
	publisher = {IOP},
	author = {Evans, Ian N. and Primini, Francis A. and Glotfelty, Kenny J. and Anderson, Craig S. and Bonaventura, Nina R. and Chen, Judy C. and Davis, John E. and Doe, Stephen M. and Evans, Janet D. and Fabbiano, Giuseppina and Galle, Elizabeth C. and Gibbs, II, Danny G. and Grier, John D. and Hain, Roger M. and Hall, Diane M. and Harbo, Peter N. and He, Xiangqun Helen and Houck, John C. and Karovska, Margarita and Kashyap, Vinay L. and Lauer, Jennifer and McCollough, Michael L. and McDowell, Jonathan C. and Miller, Joseph B. and Mitschang, Arik W. and Morgan, Douglas L. and Mossman, Amy E. and Nichols, Joy S. and Nowak, Michael A. and Plummer, David A. and Refsdal, Brian L. and Rots, Arnold H. and Siemiginowska, Aneta and Sundheim, Beth A. and Tibbetts, Michael S. and Van Stone, David W. and Winkelman, Sherry L. and Zografou, Panagoula},
	month = jul,
	year = {2010},
	note = {ADS Bibcode: 2010ApJS..189...37E},
	pages = {37--82},
}

@article{parker_astroclip_2024,
	title = {{AstroCLIP}: a cross-modal foundation model for galaxies},
	volume = {531},
	issn = {0035-8711},
	shorttitle = {{AstroCLIP}},
	url = {https://ui.adsabs.harvard.edu/abs/2024MNRAS.531.4990P},
	doi = {10.1093/mnras/stae1450},
	urldate = {2026-09-29},
	journal = {Monthly Notices of the Royal Astronomical Society},
	publisher = {OUP},
	author = {Parker, Liam and Lanusse, Francois and Golkar, Siavash and Sarra, Leopoldo and Cranmer, Miles and Bietti, Alberto and Eickenberg, Michael and Krawezik, Geraud and McCabe, Michael and Morel, Rudy and Ohana, Ruben and Pettee, Mariel and Régaldo-Saint Blancard, Bruno and Cho, Kyunghyun and Ho, Shirley and {Polymathic AI Collaboration}},
	month = jul,
	year = {2024},
	note = {ADS Bibcode: 2024MNRAS.531.4990P},
	pages = {4990--5011},
}

@article{dey_overview_2019,
	title = {Overview of the {DESI} {Legacy} {Imaging} {Surveys}},
	volume = {157},
	issn = {0004-6256},
	url = {https://ui.adsabs.harvard.edu/abs/2019AJ....157..168D},
	doi = {10.3847/1538-3881/ab089d},
	urldate = {2026-09-29},
	journal = {The Astronomical Journal},
	publisher = {IOP},
	author = {Dey, Arjun and Schlegel, David J. and Lang, Dustin and Blum, Robert and Burleigh, Kaylan and Fan, Xiaohui and Findlay, Joseph R. and Finkbeiner, Doug and Herrera, David and Juneau, Stéphanie and Landriau, Martin and Levi, Michael and McGreer, Ian and Meisner, Aaron and Myers, Adam D. and Moustakas, John and Nugent, Peter and Patej, Anna and Schlafly, Edward F. and Walker, Alistair R. and Valdes, Francisco and Weaver, Benjamin A. and Yèche, Christophe and Zou, Hu and Zhou, Xu and Abareshi, Behzad and Abbott, T. M. C. and Abolfathi, Bela and Aguilera, C. and Alam, Shadab and Allen, Lori and Alvarez, A. and Annis, James and Ansarinejad, Behzad and Aubert, Marie and Beechert, Jacqueline and Bell, Eric F. and BenZvi, Segev Y. and Beutler, Florian and Bielby, Richard M. and Bolton, Adam S. and Briceño, César and Buckley-Geer, Elizabeth J. and Butler, Karen and Calamida, Annalisa and Carlberg, Raymond G. and Carter, Paul and Casas, Ricard and Castander, Francisco J. and Choi, Yumi and Comparat, Johan and Cukanovaite, Elena and Delubac, Timothée and DeVries, Kaitlin and Dey, Sharmila and Dhungana, Govinda and Dickinson, Mark and Ding, Zhejie and Donaldson, John B. and Duan, Yutong and Duckworth, Christopher J. and Eftekharzadeh, Sarah and Eisenstein, Daniel J. and Etourneau, Thomas and Fagrelius, Parker A. and Farihi, Jay and Fitzpatrick, Mike and Font-Ribera, Andreu and Fulmer, Leah and Gänsicke, Boris T. and Gaztanaga, Enrique and George, Koshy and Gerdes, David W. and Gontcho, Satya Gontcho A. and Gorgoni, Claudio and Green, Gregory and Guy, Julien and Harmer, Diane and Hernandez, M. and Honscheid, Klaus and Huang, Lijuan Wendy and James, David J. and Jannuzi, Buell T. and Jiang, Linhua and Joyce, Richard and Karcher, Armin and Karkar, Sonia and Kehoe, Robert and Kneib, Jean-Paul and Kueter-Young, Andrea and Lan, Ting-Wen and Lauer, Tod R. and Le Guillou, Laurent and Le Van Suu, Auguste and Lee, Jae Hyeon and Lesser, Michael and Perreault Levasseur, Laurence and Li, Ting S. and Mann, Justin L. and Marshall, Robert and Martínez-Vázquez, C. E. and Martini, Paul and du Mas des Bourboux, Hélion and McManus, Sean and Meier, Tobias Gabriel and Ménard, Brice and Metcalfe, Nigel and Muñoz-Gutiérrez, Andrea and Najita, Joan and Napier, Kevin and Narayan, Gautham and Newman, Jeffrey A. and Nie, Jundan and Nord, Brian and Norman, Dara J. and Olsen, Knut A. G. and Paat, Anthony and Palanque-Delabrouille, Nathalie and Peng, Xiyan and Poppett, Claire L. and Poremba, Megan R. and Prakash, Abhishek and Rabinowitz, David and Raichoor, Anand and Rezaie, Mehdi and Robertson, A. N. and Roe, Natalie A. and Ross, Ashley J. and Ross, Nicholas P. and Rudnick, Gregory and Safonova, Sasha and Saha, Abhijit and Sánchez, F. Javier and Savary, Elodie and Schweiker, Heidi and Scott, Adam and Seo, Hee-Jong and Shan, Huanyuan and Silva, David R. and Slepian, Zachary and Soto, Christian and Sprayberry, David and Staten, Ryan and Stillman, Coley M. and Stupak, Robert J. and Summers, David L. and Sien Tie, Suk and Tirado, H. and Vargas-Magaña, Mariana and Vivas, A. Katherina and Wechsler, Risa H. and Williams, Doug and Yang, Jinyi and Yang, Qian and Yapici, Tolga and Zaritsky, Dennis and Zenteno, A. and Zhang, Kai and Zhang, Tianmeng and Zhou, Rongpu and Zhou, Zhimin},
	month = may,
	year = {2019},
	note = {ADS Bibcode: 2019AJ....157..168D},
	pages = {168},
}

@article{liu_method_2015,
	title = {A Method to Search for Bulk Motions in the {ICM} with \textit{{Chandra}} {CCD} Spectra: Application to the {Bullet} Cluster},
	volume = {809},
	copyright = {http://iopscience.iop.org/info/page/text-and-data-mining},
	issn = {1538-4357},
	shorttitle = {A {METHOD} {TO} {SEARCH} {FOR} {BULK} {MOTIONS} {IN} {THE} {ICM} {WITH} \textit{{CHANDRA}} {CCD} {SPECTRA}},
	url = {https://iopscience.iop.org/article/10.1088/0004-637X/809/1/27},
	doi = {10.1088/0004-637X/809/1/27},
	language = {en},
	number = {1},
	urldate = {2026-09-29},
	journal = {ApJ},
	author = {Liu, Ang and Yu, Heng and Tozzi, Paolo and Zhu, Zong-Hong},
	month = aug,
	year = {2015},
	pages = {27},
}

@article{shimwell_another_2015,
	title = {Another shock for the {Bullet} cluster, and the source of seed electrons for radio relics},
	volume = {449},
	issn = {0035-8711},
	url = {https://doi.org/10.1093/mnras/stv334},
	doi = {10.1093/mnras/stv334},
	number = {2},
	urldate = {2026-09-29},
	journal = {Mon Not R Astron Soc},
	author = {Shimwell, Timothy W. and Markevitch, Maxim and Brown, Shea and Feretti, Luigina and Gaensler, B. M. and Johnston-Hollitt, M. and Lage, Craig and Srinivasan, Raghav},
	month = may,
	year = {2015},
	pages = {1486--1494},
}

@article{clowe_direct_2006,
	title = {A {Direct} {Empirical} {Proof} of the {Existence} of {Dark} {Matter}},
	volume = {648},
	issn = {0004-637X},
	url = {https://ui.adsabs.harvard.edu/abs/2006ApJ...648L.109C},
	doi = {10.1086/508162},
	urldate = {2026-09-29},
	journal = {The Astrophysical Journal Letters},
	publisher = {IOP},
	author = {Clowe, Douglas and Bradač, Maruša and Gonzalez, Anthony H. and Markevitch, Maxim and Randall, Scott W. and Jones, Christine and Zaritsky, Dennis},
	month = sep,
	year = {2006},
	note = {ADS Bibcode: 2006ApJ...648L.109C},
	pages = {L109--L113},
}

@article{markevitch_direct_2004,
	title = {Direct {Constraints} on the {Dark} {Matter} {Self}-{Interaction} {Cross} {Section} from the {Merging} {Galaxy} {Cluster} {1E} 0657-56},
	volume = {606},
	issn = {0004-637X},
	url = {https://ui.adsabs.harvard.edu/abs/2004ApJ...606..819M},
	doi = {10.1086/383178},
	urldate = {2026-09-29},
	journal = {The Astrophysical Journal},
	publisher = {IOP},
	author = {Markevitch, M. and Gonzalez, A. H. and Clowe, D. and Vikhlinin, A. and Forman, W. and Jones, C. and Murray, S. and Tucker, W.},
	month = may,
	year = {2004},
	note = {ADS Bibcode: 2004ApJ...606..819M},
	pages = {819--824},
}

@article{markevitch_textbook_2002,
	title = {A {Textbook} {Example} of a {Bow} {Shock} in the {Merging} {Galaxy} {Cluster} {1E} 0657-56},
	volume = {567},
	issn = {0004-637X},
	url = {https://ui.adsabs.harvard.edu/abs/2002ApJ...567L..27M},
	doi = {10.1086/339619},
	urldate = {2026-09-29},
	journal = {The Astrophysical Journal Letters},
	publisher = {IOP},
	author = {Markevitch, M. and Gonzalez, A. H. and David, L. and Vikhlinin, A. and Murray, S. and Forman, W. and Jones, C. and Tucker, W.},
	month = mar,
	year = {2002},
	note = {ADS Bibcode: 2002ApJ...567L..27M},
	pages = {L27--L31},
}

@article{kim_chandra_2019,
	title = {Chandra {Early}-type {Galaxy} {Atlas}},
	volume = {241},
	issn = {0067-0049},
	url = {https://doi.org/10.3847/1538-4365/ab0ca4},
	doi = {10.3847/1538-4365/ab0ca4},
	language = {en},
	number = {2},
	urldate = {2026-09-29},
	journal = {ApJS},
	publisher = {The American Astronomical Society},
	author = {Kim, Dong-Woo and Anderson, Craig and Burke, Douglas and D’Abrusco, Raffaele and Fabbiano, Giuseppina and Fruscione, Antonella and Lauer, Jennifer and McCollough, Michael and Morgan, Douglas and Mossman, Amy and O’Sullivan, Ewan and Paggi, Alessandro and Vrtilek, Saeqa and Trinchieri, Ginevra},
	month = apr,
	year = {2019},
	pages = {36},
}

@article{posson-brown_chandra_2009,
	title = {{Chandra} Observations of the {X}-ray Point Source Population in {NGC} 4636},
	volume = {695},
	issn = {0004-637X},
	url = {https://doi.org/10.1088/0004-637X/695/2/1094},
	doi = {10.1088/0004-637X/695/2/1094},
	language = {en},
	number = {2},
	urldate = {2026-09-29},
	journal = {ApJ},
	publisher = {The American Astronomical Society},
	author = {Posson-Brown, Jennifer and Raychaudhury, Somak and Forman, William and Donnelly, R. Hank and Jones, Christine},
	month = apr,
	year = {2009},
	pages = {1094},
}

@article{baldi_unusual_2009,
	title = {The Unusual {X}-ray Morphology of {NGC} 4636 Revealed by Deep {Chandra} Observations: Cavities and Shocks Created by Past Active Galactic Nucleus Outbursts},
	volume = {707},
	issn = {0004-637X},
	shorttitle = {{THE} {UNUSUAL} {X}-{RAY} {MORPHOLOGY} {OF} {NGC} 4636 {REVEALED} {BY} {DEEP} {CHANDRA} {OBSERVATIONS}},
	url = {https://doi.org/10.1088/0004-637X/707/2/1034},
	doi = {10.1088/0004-637X/707/2/1034},
	language = {en},
	number = {2},
	urldate = {2026-09-29},
	journal = {ApJ},
	publisher = {The American Astronomical Society},
	author = {Baldi, A. and Forman, W. and Jones, C. and Kraft, R. and Nulsen, P. and Churazov, E. and David, L. and Giacintucci, S.},
	month = dec,
	year = {2009},
	pages = {1034},
}

@article{osullivan_active_2005,
	title = {Active {Galactic} {Nucleus} {Feedback} and {Gas} {Mixing} in the {Core} of {NGC} 4636},
	volume = {624},
	issn = {0004-637X},
	url = {https://ui.adsabs.harvard.edu/abs/2005ApJ...624L..77O},
	doi = {10.1086/430600},
	urldate = {2026-09-29},
	journal = {The Astrophysical Journal Letters},
	publisher = {IOP},
	author = {O'Sullivan, E. and Vrtilek, J. M. and Kempner, J. C.},
	month = may,
	year = {2005},
	note = {ADS Bibcode: 2005ApJ...624L..77O},
	pages = {L77--L80},
}

@article{jones_chandra_2002,
	title = {\textit{{Chandra}} {Observations} of {NGC} 4636--an {Elliptical} {Galaxy} in {Turmoil}},
	volume = {567},
	issn = {0004-637X, 1538-4357},
	url = {https://iopscience.iop.org/article/10.1086/340114},
	doi = {10.1086/340114},
	language = {en},
	number = {2},
	urldate = {2026-09-29},
	journal = {ApJ},
	author = {Jones, C. and Forman, W. and Vikhlinin, A. and Markevitch, M. and David, L. and Warmflash, A. and Murray, S. and Nulsen, P. E. J.},
	month = mar,
	year = {2002},
	pages = {L115--L118},
}

@article{donahue_accept_2026,
	title = {{ACCEPT} 2.0: {A} {Compilation} of {X}-{Ray} {Gas} {Properties} for {Clusters} of {Galaxies} {Observed} by the {Chandra} {X}-{Ray} {Observatory}},
	volume = {282},
	issn = {0067-0049},
	shorttitle = {{ACCEPT} 2.0},
	url = {https://ui.adsabs.harvard.edu/abs/2026ApJS..282...61D},
	doi = {10.3847/1538-4365/ae2c5f},
	urldate = {2026-10-16},
	journal = {The Astrophysical Journal Supplement Series},
	publisher = {IOP},
	author = {Donahue, Megan and Koeppe, Dana and Frisbie, Rachel and Baldi, Allesandro and Voit, G. Mark},
	month = feb,
	year = {2026},
	note = {ADS Bibcode: 2026ApJS..282...61D},
	pages = {61},
}

@article{harris_array_2020,
	title = {Array programming with {NumPy}},
	volume = {585},
	copyright = {2020 The Author(s)},
	issn = {1476-4687},
	url = {https://www.nature.com/articles/s41586-020-2649-2},
	doi = {10.1038/s41586-020-2649-2},
	language = {en},
	number = {7825},
	urldate = {2026-10-24},
	journal = {Nature},
	publisher = {Nature Publishing Group},
	author = {Harris, Charles R. and Millman, K. Jarrod and van der Walt, Stéfan J. and Gommers, Ralf and Virtanen, Pauli and Cournapeau, David and Wieser, Eric and Taylor, Julian and Berg, Sebastian and Smith, Nathaniel J. and Kern, Robert and Picus, Matti and Hoyer, Stephan and van Kerkwijk, Marten H. and Brett, Matthew and Haldane, Allan and del Río, Jaime Fernández and Wiebe, Mark and Peterson, Pearu and Gérard-Marchant, Pierre and Sheppard, Kevin and Reddy, Tyler and Weckesser, Warren and Abbasi, Hameer and Gohlke, Christoph and Oliphant, Travis E.},
	month = sep,
	year = {2020},
	pages = {357--362},
}

@misc{karam_cool_2026,
	title = {Cool {Embeddings}: {Predicting} {Galaxy} {Cluster} {Cooling} {Times} in {IllustrisTNG} and {TNG}-{Cluster} with {AstroCLIP}},
	shorttitle = {Cool {Embeddings}},
	url = {https://ui.adsabs.harvard.edu/abs/2026arXiv260932582K},
	doi = {10.48550/arXiv.2609.32582},
	urldate = {2026-10-20},
	publisher = {arXiv},
	author = {Karam, Rawan and Rhea, Carter Lee and Hlavacek-Larrondo, Julie and Perreault-Levasseur, Laurence and Prunier, Marine and Alexandre, Adam and Choi, Hyunseop},
	month = sep,
	year = {2026},
	note = {ADS Bibcode: 2026arXiv260932582K},
}

@article{virtanen_scipy_2020,
	title = {{SciPy} 1.0: fundamental algorithms for scientific computing in {Python}},
	volume = {17},
	copyright = {2020 The Author(s)},
	issn = {1548-7105},
	shorttitle = {{SciPy} 1.0},
	url = {https://www.nature.com/articles/s41592-019-0686-2},
	doi = {10.1038/s41592-019-0686-2},
	language = {en},
	number = {3},
	urldate = {2026-10-24},
	journal = {Nat Methods},
	publisher = {Nature Publishing Group},
	author = {Virtanen, Pauli and Gommers, Ralf and Oliphant, Travis E. and Haberland, Matt and Reddy, Tyler and Cournapeau, David and Burovski, Evgeni and Peterson, Pearu and Weckesser, Warren and Bright, Jonathan and van der Walt, Stéfan J. and Brett, Matthew and Wilson, Joshua and Millman, K. Jarrod and Mayorov, Nikolay and Nelson, Andrew R. J. and Jones, Eric and Kern, Robert and Larson, Eric and Carey, C. J. and Polat, İlhan and Feng, Yu and Moore, Eric W. and VanderPlas, Jake and Laxalde, Denis and Perktold, Josef and Cimrman, Robert and Henriksen, Ian and Quintero, E. A. and Harris, Charles R. and Archibald, Anne M. and Ribeiro, Antônio H. and Pedregosa, Fabian and van Mulbregt, Paul},
	month = mar,
	year = {2020},
	pages = {261--272},
}

@article{fruscione_ciao_2026,
	title = {{CIAO}: {Chandra}’s {Data} {Analysis} {System} for {X}-{Ray} {Astronomy} and {Beyond}},
	volume = {1005},
	issn = {0004-637X},
	shorttitle = {{CIAO}},
	url = {https://doi.org/10.3847/1538-4357/ae6db4},
	doi = {10.3847/1538-4357/ae6db4},
	language = {en},
	number = {1},
	urldate = {2026-10-24},
	journal = {ApJ},
	publisher = {The American Astronomical Society},
	author = {Fruscione, Antonella and McDowell, Jonathan C. and Burke, Douglas J. and Cresitello-Dittmar, Mark and Evans, Ian N. and Evans, Janet D. and Glotfelty, Kenny and Günther, Hans Moritz and Huenemoerder, David and Joye, William and Lee, Nicholas P. and McLaughlin, Warren and Miller, Joseph B. and Nynka, Melania and Principe, David A. and Siemiginowska, Aneta},
	month = jun,
	year = {2026},
	pages = {116},
}

@article{collaboration_astropy_2022,
	title = {The {Astropy} {Project}: {Sustaining} and {Growing} a {Community}-oriented {Open}-source {Project} and the {Latest} {Major} {Release} (v5.0) of the {Core} {Package}*},
	volume = {935},
	issn = {0004-637X},
	shorttitle = {The {Astropy} {Project}},
	url = {https://doi.org/10.3847/1538-4357/ac7c74},
	doi = {10.3847/1538-4357/ac7c74},
	language = {en},
	number = {2},
	urldate = {2026-10-24},
	journal = {ApJ},
	publisher = {The American Astronomical Society},
	author = {{Astropy Collaboration} and Price-Whelan, Adrian M. and Lim, Pey Lian and Earl, Nicholas and Starkman, Nathaniel and Bradley, Larry and Shupe, David L. and Patil, Aarya A. and Corrales, Lia and Brasseur, C. E. and Nöthe, Maximilian and Donath, Axel and Tollerud, Erik and Morris, Brett M. and Ginsburg, Adam and Vaher, Eero and Weaver, Benjamin A. and Tocknell, James and Jamieson, William and van Kerkwijk, Marten H. and Robitaille, Thomas P. and Merry, Bruce and Bachetti, Matteo and Günther, H. Moritz and Aldcroft, Thomas L. and Alvarado-Montes, Jaime A. and Archibald, Anne M. and Bódi, Attila and Bapat, Shreyas and Barentsen, Geert and Bazán, Juanjo and Biswas, Manish and Boquien, Médéric and Burke, D. J. and Cara, Daria and Cara, Mihai and Conroy, Kyle E and Conseil, Simon and Craig, Matthew W. and Cross, Robert M. and Cruz, Kelle L. and D’Eugenio, Francesco and Dencheva, Nadia and Devillepoix, Hadrien A. R. and Dietrich, Jörg P. and Eigenbrot, Arthur Davis and Erben, Thomas and Ferreira, Leonardo and Foreman-Mackey, Daniel and Fox, Ryan and Freij, Nabil and Garg, Suyog and Geda, Robel and Glattly, Lauren and Gondhalekar, Yash and Gordon, Karl D. and Grant, David and Greenfield, Perry and Groener, Austen M. and Guest, Steve and Gurovich, Sebastian and Handberg, Rasmus and Hart, Akeem and Hatfield-Dodds, Zac and Homeier, Derek and Hosseinzadeh, Griffin and Jenness, Tim and Jones, Craig K. and Joseph, Prajwel and Kalmbach, J. Bryce and Karamehmetoglu, Emir and Kałuszyński, Mikołaj and Kelley, Michael S. P. and Kern, Nicholas and Kerzendorf, Wolfgang E. and Koch, Eric W. and Kulumani, Shankar and Lee, Antony and Ly, Chun and Ma, Zhiyuan and MacBride, Conor and Maljaars, Jakob M. and Muna, Demitri and Murphy, N. A. and Norman, Henrik and O’Steen, Richard and Oman, Kyle A. and Pacifici, Camilla and Pascual, Sergio and Pascual-Granado, J. and Patil, Rohit R. and Perren, Gabriel I and Pickering, Timothy E. and Rastogi, Tanuj and Roulston, Benjamin R. and Ryan, Daniel F and Rykoff, Eli S. and Sabater, Jose and Sakurikar, Parikshit and Salgado, Jesús and Sanghi, Aniket and Saunders, Nicholas and Savchenko, Volodymyr and Schwardt, Ludwig and Seifert-Eckert, Michael and Shih, Albert Y. and Jain, Anany Shrey and Shukla, Gyanendra and Sick, Jonathan and Simpson, Chris and Singanamalla, Sudheesh and Singer, Leo P. and Singhal, Jaladh and Sinha, Manodeep and Sipőcz, Brigitta M. and Spitler, Lee R. and Stansby, David and Streicher, Ole and Šumak, Jani and Swinbank, John D. and Taranu, Dan S. and Tewary, Nikita and Tremblay, Grant R. and Val-Borro, Miguel de and Van Kooten, Samuel J. and Vasović, Zlatan and Verma, Shresth and de Miranda Cardoso, José Vinícius and Williams, Peter K. G. and Wilson, Tom J. and Winkel, Benjamin and Wood-Vasey, W. M. and Xue, Rui and Yoachim, Peter and Zhang, Chen and Zonca, Andrea},
	month = aug,
	year = {2022},
	pages = {167},
}

@book{van_rossum_python_2009,
	address = {Scotts Valley, CA},
	title = {Python 3 {Reference} {Manual}},
	isbn = {978-1-4414-1269-0},
	publisher = {CreateSpace},
	author = {Van Rossum, Guido and Drake, Fred L.},
	month = feb,
	year = {2009},
}

@inproceedings{mckinney_data_2010,
	booktitle = {Proceedings of the 9th {Python} in {Science} {Conference}},
	address = {Austin, Texas},
	title = {Data {Structures} for {Statistical} {Computing} in {Python}},
	url = {https://doi.curvenote.com/10.25080/Majora-92bf1922-00a},
	doi = {10.25080/Majora-92bf1922-00a},
	language = {en},
	urldate = {2026-10-24},
	author = {McKinney, Wes},
	year = {2010},
	pages = {56--61},
}

@misc{ramirez_fastapi_2018,
	title = {{FastAPI}},
	url = {https://github.com/fastapi/fastapi},
	author = {Ramirez, Sebastien},
	year = {2018},
}

@inproceedings{fruscione_ciao_2006,
	title = {{CIAO}: {Chandra}'s data analysis system},
	volume = {6270},
	doi = {10.1117/12.671760},
	booktitle = {Observatory {Operations}: {Strategies}, {Processes}, and {Systems}},
	series = {Society of {Photo}-{Optical} {Instrumentation} {Engineers} ({SPIE}) {Conference} {Series}},
	editor = {Silva, David R. and Doxsey, Rodger E.},
	author = {Fruscione, Antonella and McDowell, Jonathan C. and Allen, Glenn E. and Brickhouse, Nancy S. and Burke, Douglas J. and Davis, John E. and Durham, Nick and Elvis, Martin and Galle, Elizabeth C. and Harris, Daniel E. and Huenemoerder, David P. and Houck, John C. and Ishibashi, Bish and Karovska, Margarita and Nicastro, Fabrizio and Noble, Michael S. and Nowak, Michael A. and Primini, Frank A. and Siemiginowska, Aneta and Smith, Randall K. and Wise, Michael},
	month = jun,
	year = {2006},
	note = {ADS Bibcode: 2006SPIE.6270E..1VF},
	pages = {62701V},
}

@inproceedings{markevitch_2006,
	title = {Chandra observation of the most interesting cluster in the {Universe}},
	volume = {604},
	booktitle = {The {X}-ray {Universe} 2005},
	series = {ESA Special Publication},
	editor = {Wilson, A.},
	author = {Markevitch, Maxim},
	month = jan,
	year = {2006},
	note = {ADS Bibcode: 2006ESASP.604..723M; arXiv:astro-ph/0511345},
	pages = {723},
}

@software{pandas_development_team_2026,
	author = {{The pandas development team}},
	title = {pandas-dev/pandas: Pandas},
	version = {v3.1.0},
	publisher = {Zenodo},
	year = {2026},
	doi = {10.5281/zenodo.23565656},
	url = {https://doi.org/10.5281/zenodo.23565656},
}

@article{wenger_simbad_2000,
	author = {Wenger, M. and Ochsenbein, F. and Egret, D. and Dubois, P. and Bonnarel, F. and Borde, S. and Genova, F. and Jasniewicz, G. and Lalo{\"e}, S. and Lesteven, S. and Monier, R.},
	title = {The {SIMBAD} astronomical database. {The} {CDS} reference database for astronomical objects},
	journal = {Astronomy and Astrophysics Supplement Series},
	volume = {143},
	pages = {9--22},
	year = {2000},
	doi = {10.1051/aas:2000332},
}

@article{planck_collaboration_planck_2020,
	author = {{Planck Collaboration} and Aghanim, N. and Akrami, Y. and Ashdown, M. and others},
	title = {Planck 2018 results. {VI}. {Cosmological} parameters},
	journal = {Astronomy and Astrophysics},
	volume = {641},
	pages = {A6},
	year = {2020},
	doi = {10.1051/0004-6361/201833910},
}

@inproceedings{plucinsky_complicated_2018,
	author = {Plucinsky, Paul P. and Bogdan, Akos and Marshall, Herman L. and Tice, Neil W.},
	title = {The complicated evolution of the {ACIS} contamination layer over the mission life of the {Chandra} {X}-ray {Observatory}},
	booktitle = {Space Telescopes and Instrumentation 2018: Ultraviolet to Gamma Ray},
	series = {Society of Photo-Optical Instrumentation Engineers (SPIE) Conference Series},
	volume = {10699},
	year = {2018},
	eprint = {1809.02225},
	archivePrefix = {arXiv},
}



\end{document}